\documentclass[aps,pre,reprint,groupedaddress]{revtex4-1}

\pdfoutput=1
\usepackage{graphicx,amsmath}

\usepackage{xcolor}
\usepackage{braket}
\usepackage{xr}

\begin{document}


\title{Ranking nodes in growing networks: When PageRank fails}


\author{Manuel Sebastian Mariani, Mat{\'u}{\v{s}} Medo, \& Yi-Cheng Zhang}
\affiliation{Department of Physics, University of Fribourg, 1700 Fribourg, Switzerland}



\begin{abstract}
PageRank is arguably the most popular ranking algorithm which
is being applied in real systems ranging from information
to biological and infrastructure networks.
Despite its outstanding popularity and broad use in different areas of science, the relation between
the algorithm's efficacy and properties of the network on which it acts has not yet been fully understood.
We study here PageRank's performance on a network model supported by real data, 
and show that realistic temporal effects make PageRank fail in individuating the most valuable nodes for a broad range of model parameters.
Results on real data are in qualitative agreement with our model-based
findings.
This failure of PageRank reveals that the 
static approach to information filtering
is inappropriate for a broad class of growing systems, and suggest that time-dependent algorithms 
that are based on the temporal linking patterns of these systems are needed to
better rank the nodes.
\end{abstract}


\maketitle



\section{Introduction}

With the amount of available information constantly growing due to the widespread usage of computers and the Internet, 
network-driven information filtering tools such as 
ranking algorithms~\cite{duhan2009page,medo2013network} and 
recommender systems~\cite{lu2012recommender} 
attract attention of researchers from various fields.
PageRank, one of the most popular ranking algorithms, has been originally devised to rank web sites in search engine results~\cite{brin1998anatomy}. 
The algorithm acts on unipartite directed networks and builds on the circular idea \emph{``A node is important if it is pointed by other important nodes''}. 
The essential role that PageRank plays in the Google search algorithm has stimulated extensive research of its properties~\cite{langville2004deeper}
and relations to previous ranking techniques~\cite{franceschet2011pagerank}. 
PageRank has been applied far beyond its original scope: in
ranking of scholarly papers~\cite{chen2007finding}, authors~\cite{ding2009pagerank,yan2011discovering} and journals~\cite{bollen2006journal}, ranking of images 
in search~\cite{jing2008visualrank}, ranking of urban roads according to traffic flow \cite{jiang2008self}, 
measuring the importance of biochemical reactions in the metabolic network~\cite{ivan2011web}, for example.
The algorithm's remarkable stability properties~\cite{langville2004deeper, ghoshal2011ranking} make it a suitable candidate to rank nodes
in noisy networks such as the World Wide Web (WWW) and the protein interaction networks, where the information is often not completely reliable.
Variants of PageRank include Eigentrust which computes trust values in distributed peer-to-peer systems~\cite{kamvar2003eigentrust}, LeaderRank which computes 
influence of users in social networks~\cite{lu2011leaders}, 
and CiteRank which uses a model of citation network traffic to compute the importance of 
scientific papers~\cite{walker2007ranking}, among others; variants of PageRank have been also applied to bipartite 
networks \cite{allesina2009googling,tacchella2012new,dominguez2015ranking}
and multilayer networks \cite{de2015ranking}.

The widespread usage of PageRank motivates us to 
ask: when is the algorithm effective in ranking nodes according to their quality?
Are there circumstances under which the algorithm is doomed to fail?
Answering these questions is of primary importance to foster our understanding of the ranking algorithm,
which is a problem of practical significance given the influence of ranking-based tools such as search engines and recommendation systems
on many aspects of our society, from marketing to politics \cite{hindman2003googlearchy,cho2004impact,anderson2006long,fortunato2006topical}.
While previous research has already studied the rankings produced by PageRank for different 
topological properties of the input networks \cite{ghoshal2011ranking},
the evaluation of the algorithm on networks that evolve in time remains a largely unexplored field. 
The main aim of this work is to fill this gap and demonstrate the shortcomings of the algorithm when applied to growing networks exhibiting temporal effects.
To this end, we use a growing directed network model with
preferential attachment and relevance~\cite{medo2011temporal}
which generalizes the classical preferential attachment introduced in~\cite{barabasi1999emergence}.
This model (hereafter the Relevance Model, RM) has been shown by maximum likelihood analysis
to be the preferential attachment model that best explains 
the linking patterns in real information systems~\cite{medo2014statistical} and has been used to model real information systems, such as
the WWW~\cite{kong2008experience}, citation networks~\cite{wang2013quantifying}, online networks \cite{medo2014statistical},
and even technological networks, such as the
network of Internet autonomous systems~\cite{vazquez2002large}.

In the RM, three essential elements rule the competition among nodes for incoming links: preferential attachment, fitness and temporal decay.
Preferential attachment is a well-established mechanism that has been observed in a wide range of
real systems (see~\cite{newman2010networks, dorogovtsev2013evolution} for a review).
\emph{Fitness} is a quality parameter assigned to each node that quantifies
the node's inherent competence in attracting new incoming
links~\cite{bianconi2001competition}; all other things being equal, in a competitive environment high-fitness nodes are suitable for success in the system and
are likely to become eventually popular, whereas low fitness nodes tend to remain little known \cite{kong2008experience}. 
Node fitness is modulated with a time-decaying function
which gives rise to the so-called node relevance \cite{medo2011temporal}:
a node of high-fitness thus initially has high relevance and potentially attracts many links but this relevance eventually vanishes and the node ceases
to attract new links.
Fitness and relevance discount all system-dependent intangible and subjective factors that determine node's quality,
quantify how much a node is attractive to a given system and can be estimated on real
data by different techniques~\cite{kong2008experience,medo2011temporal,wang2013quantifying, medo2014statistical}. 
In our model, each node is further endowed with an activity parameter which represents the rate at which the node creates
new outgoing links;
activity too is modulated with time.
We
use the model to produce artificial data and compare the ranking of nodes by their indegree (i.e., the number of incoming links)
and PageRank score with the node ranking by their fitness values.
We find that when model parameters for the temporal decay of relevance and activity substantially differ from each other, 
the redistribution of PageRank scores is biased towards old or recent nodes, respectively (depending on which decay is faster). In addition, 
when PageRank is temporally biased in either way, indegree markedly outperforms it in ranking nodes by their fitness.
These results are confirmed
on a modified model, so-called Extended Fitness Model, 
where high-fitness nodes preferentially attach to other high-fitness nodes, 
whereas low-fitness nodes preferentially attach to popular nodes. While in this model PageRank 
can significantly outperform indegree in reproducing the ranking of nodes by their fitness for some model parameters,
extensive parameter regions where the algorithm fails and performs worse than indegree are still present.

We finally apply PageRank on two real datasets, the social network of Digg.com users
and the network of citations between American Physical Society (APS) scientific articles, 
and compare the rankings of nodes by their indegree and PageRank score with the node ranking by their total relevance which is a real-data estimate for fitness.
We find
that while PageRank score is highly correlated with indegree in social network data and the two metrics have similar performance,
PageRank is markedly outperformed by indegree in citation data.
These findings strongly discourage the use of PageRank in systems where strong temporal patterns exist, like citation networks.

\section{Results}

 \begin{figure*}[t]
 \centering
 \includegraphics[width=1.5\columnwidth, angle=0]{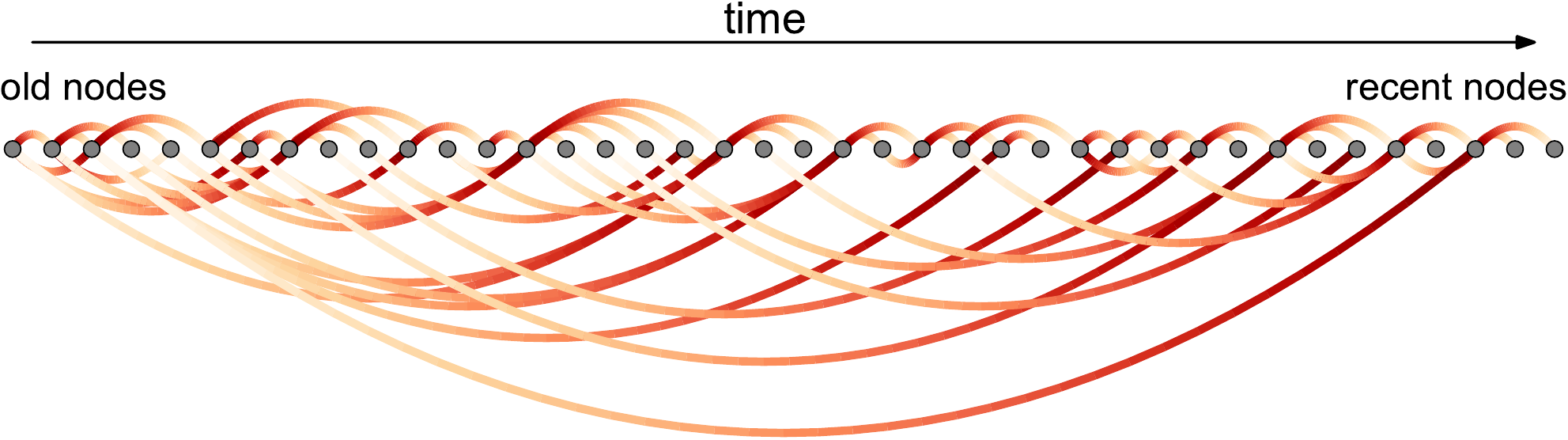}
 \caption{\textbf{Illustration of a network produced with the RM for fast aging of relevance and constant activity} (color online).
 In each step, a new node is introduced and connected to an existing node (arcs above the row of nodes).
 In addition, a randomly chosen node becomes active and connects to an existing node (arcs below). 
 The target node is chosen by Eq. \eqref{PARD} in both cases (see Supplementary Note S5 for model parameters).
 The orange and red part of the each link mark the initial and target node, respectively.
 Note that while old nodes point to nodes of every age thanks to constant activity, 
 recent nodes never point to the old nodes due to the decay of relevance. This asymmetry results in PageRank scores biased towards recent nodes.
 \label{fig:bias1}}
\end{figure*}

\begin{figure}[t]
\centering
\includegraphics[height=0.8\columnwidth,angle=270]{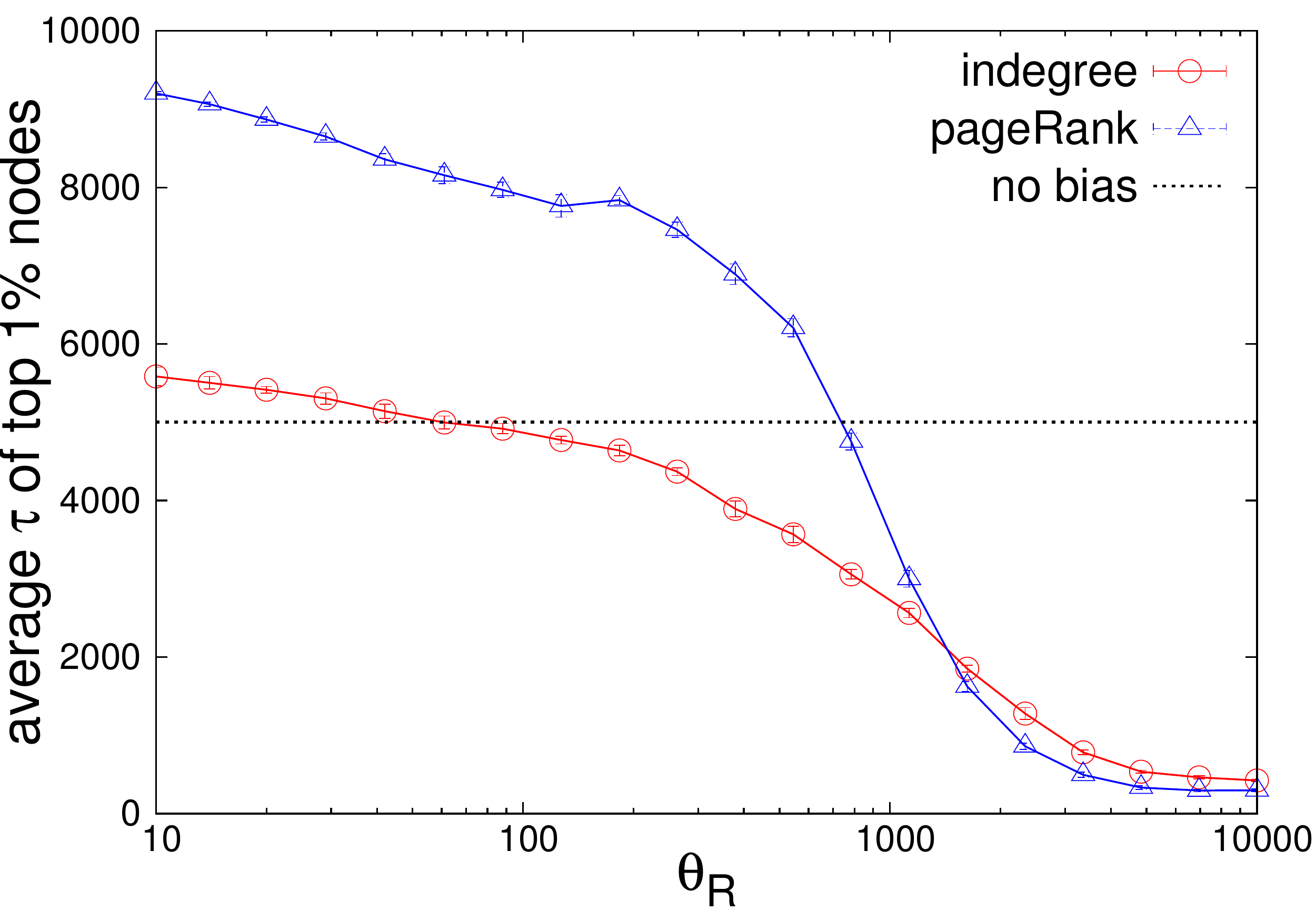}
\caption{\textbf{PageRank time bias} (color online). We show here the average entrance time $\tau$
of the top $1\%$ nodes of the node ranking
by indegree and PageRank, respectively,
as a function of the relevance decay parameter $\theta_{R}$. Networks of $N=10000$ nodes are grown with the RM with
slow decay of activity ($\theta_{A}=N$). 
Two limits of PageRank bias are visible: (1) When the decay of relevance is fast ($\theta_R \ll \theta_A$),
a large number of top nodes are recent as a consequence of the network structure demonstrated in Fig. \ref{fig:bias1};
(2) When the decay of relevance is slow ($\theta_{R}\sim N$), top nodes are old because the old nodes can be pointed by nodes of every age.
While the latter bias is common to PageRank and indegree, the former bias is specific to PageRank because of its network nature.\label{fig:bias2}}
\end{figure}

\subsection{Relevance Model (RM)}
In the RM, when a node $j$ creates a new link at time $t$, 
the probability $\Pi_{i}^{in}(t)$ 
that it chooses node $i$ as the target is assumed to be
\begin{equation}
 \Pi_{i}^{in}(t)\sim (k_{i}^{in}(t)+1)\,\eta_{i}\,f_{R}(t-\tau_{i})
 \label{PARD}
\end{equation}
where $k_{i}^{in}(t)$ is the current indegree of node $i$, $\eta_{i}$ is its fitness and $f_{R}$ 
is a function of the node's age ($\tau_{i}$ is the time at which node $i$ enters the system).
The product $\eta_{i}\,f_{R}(t-\tau_{i}):=R_{i}(t)$ represents the \emph{relevance} of node $i$ at time $t$~\cite{medo2011temporal, medo2014statistical}.
We assume that $f_{R}(t)$ decays monotonously and thus mimics real situations where nodes lose relevance over time. 
Previous studies of the RM \cite{medo2011temporal, wang2013quantifying} have focused on scientific citation networks which are tree-like because
nodes create outgoing links only in the moment when they enter the system -- the links are thus always directed back in time.
We consider a general situation where nodes continue being active, create outgoing links continually, and the resulting network thus contains loops which
are common in many real systems, such as the WWW, for example.
We use the
activity potential approach introduced in~\cite{perra2012activity}
and assign to each node $i$ an activity parameter $A_{i}$.
At each simulation step, a new node is created and connected to an existing node. In addition, $m=10$ existing nodes
are sequentially chosen and create one link each (see the Methods section for all simulation details).
The $m$ nodes that are active at time $t$ are chosen with the probability
\begin{equation}
 \Pi_{i}^{out}(t)\sim A_{i}\,f_{A}(t-\tau_{i})
 \label{activity}
\end{equation}
where $f_{A}(t)$ is a monotonously decaying function of time.
A broad distribution of the activity parameter $A$ allows us to reproduce broad outdegree distributions typically found
in real networks \cite{dorogovtsev2013evolution} without resorting
to preferential linking mechanisms for outgoing links.

\subsection{Decay of empirical relevance and activity in real data} We now analyze real data to validate the hypothesis of relevance and activity decay.
We refrain from maximum likelihood analysis \cite{medo2014statistical} because of its computational complexity.
Instead, we follow a simpler procedure: following \cite{medo2011temporal}, we define the empirical relevance
$r_{i}(t)$ of node $i$ at time $t$ as
\begin{equation}
 r_{i}(t)=\frac{n_{i}(t)}{n^{PA}_{i}(t)}.
 \label{empiricalrelevance}
\end{equation}
Here $n_{i}(t)=\Delta k^{in}_{i}(t,\Delta t)/L(t,\Delta t)$ is the ratio between the number $\Delta k^{in}_{i}(t,\Delta t)$ of incoming links received 
by node $i$ in a suitable time window $[t,t+\Delta t]$ and the total number $L(t,\Delta t)$ of links created within the same time window,
whereas $n^{PA}_{i}(t)=k^{in}_{i}(t)/\sum_{j}k^{in}_{j}(t)$ is the expected value of $n_{i}(t)$ according to preferential attachment alone.
Empirical relevance $r_{i}(t)$ larger or smaller than one means that node $i$ at time $t$ outperforms or underperforms, respectively,
with respect to its
preferential attachment weight $n^{PA}_{i}(t)$ in the competition for incoming links.

The hypothesis of time-dependent and heterogeneous relevance has already 
been validated in the APS scientific citation network~\cite{medo2011temporal}.
Here we further analyze the APS dataset, described in the Methods section, finding (Fig. S2)
that the decay of relevance is well reproduced by a power law function 
(see the Supplementary Note S2 for detailed results).
Moreover, we validate the hypothesis of relevance and activity time decay
in a very different system, the Digg.com social network of users, where a directed
link between two users means that one user follows the other (see the Methods section for the description of the dataset).
We find (Fig.S1) that relevance decays also in this dataset.
Based on \cite{perra2012activity},
we define the empirical activity $a_{i}(t)$ of node $i$ at time $t$ as the ratio between the number of outgoing links created 
by node $i$ in a suitable time window $[t,t+\Delta t]$ and the total number of links created within the same time window.
We find (Fig. S1) that also activity
decays with time, and activity decay is slower than relevance decay (see Supplementary Note S1 for details).

\subsection{Results of numerical simulation with the RM}

For the sake of generality, we consider both exponential and power-law 
decay functions $f_{R}(t)=\exp{(-t/\theta_{R})}$, $f_{A}(t)=\exp{(-t/\theta_{A})}$ and 
$f_{R}(t)=t^{-\alpha_{R}}$, $f_{A}(t)=t^{-\alpha_{A}}$, respectively.
Our main goal now is to study the dependence of PageRank performance on model parameters $\theta_{R},\theta_{A}$ and $\alpha_{R},\alpha_{A}$, respectively.
We refer to the Methods
section for the mathematical definition of PageRank and details about the choice of fitness and activity distributions in simulations.

A good ranking algorithm is expected to produce an unbiased ranking 
where both recent and old nodes have the same chance to appear at the top.
In growing networks with temporal effects, PageRank can fail to achieve this.
To explain the origin of this failure, we consider two extreme situations: 
relevance decay which is very fast and slow, respectively, with respect to activity decay.
When relevance decay is slow (or absent, as in the original fitness model \cite{bianconi2001competition}),
recent nodes receive few links because their weight in preferential attachment is much smaller than the weight of all nodes that 
have already accumulated many links
(this manifests itself in the network's strong dependence on the initial configuration \cite{berset2013effect}).
PageRank as well as indegree are therefore strongly biased towards old nodes.
When relevance decay is fast, preferential attachment is compensated by a quick decay of relevance and therefore recent nodes
can reach high indegree.
However, there is now an essential asymmetry in the system which relates to outgoing links: while
recent nodes mostly point to other recent nodes because of relevance decay,
old nodes point to nodes of every age because they remain active during the whole system's lifetime (see Fig. \ref{fig:bias1} for an illustration).
PageRank is consequently biased towards recent nodes: while a random surfer at an old node is likely to jump to a recent node,
the converse is not true; recent nodes effectively act as an attractor.

Fig. \ref{fig:bias2} shows a transition between the two extreme cases for artificial data produced by the RM with exponential relevance decay
and exponentially distributed fitness.
When the decay of relevance is slow ($\theta_R=10000$), there are only old nodes at the top $1\%$ positions of the rankings by PageRank score and indegree.
When the decay of relevance is fast ($\theta_R=10$), recent nodes occupy the majority of the top $1\%$ positions in the ranking by PageRank score.
By contrast, the ranking by indegree is essentially unbiased in this limit as the average entrance
time $\tau$ of the top-$1\%$ nodes is close to $N/2=5000$
which corresponds to the absence of time bias.

\begin{figure}[t]
\centering
\includegraphics[width=0.8\columnwidth, angle=0]{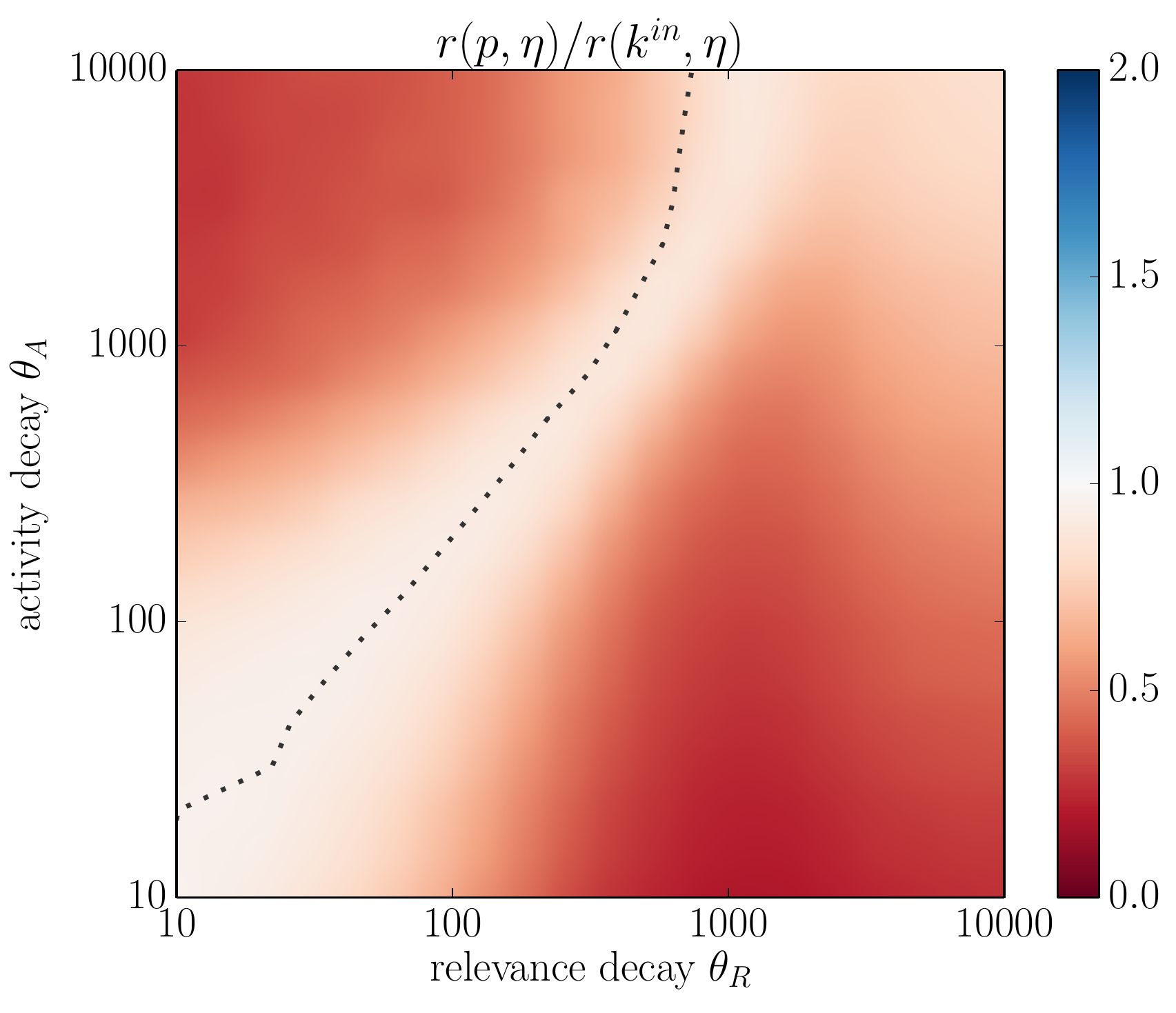}
\caption{\textbf{A comparison of performance of PageRank and indegree in the RM data} (N=10,000, $\rho(\eta)=\exp{(-\eta)}$, color online).
The heatmap shows the ratio $r(p,\eta)/r(k^{in},\eta)$. The black dotted line represents the contour along which PageRank is not temporally biased 
(see Fig. S6, left). The upward bending of this contour
is a finite-size effect. \label{fig:RM}}
\end{figure}

We discuss now the implication of PageRank's time bias on the algorithm's ability to rank nodes by fitness.
In the following, we denote by 
$r(p,\eta)$ the Pearson's correlation between the PageRank scores $p$
and the fitness values $\eta$, and we denote by $r(k^{in},\eta)$ the Pearson's correlation between node indegree and fitness.
Fig. \ref{fig:RM} shows the performance ratio $r(p,\eta)/r(k^{in},\eta)$ in the
$(\theta_{R},\theta_{A})$ plane. 
Since $r(p,\eta)/r(k^{in},\eta)<1$ everywhere, we find that PageRank yields no
improvement with respect to indegree in ranking nodes by fitness.
This is because while the
PageRank algorithm assumes that important nodes point to other important nodes, this feature is absent in the RM where
all nodes are driven by the same mechanism, Eq. \eqref{PARD}, when choosing their connections.
As a result, PageRank does best in comparison with indegree along the $\theta_{R}-\theta_{A}$ diagonal
 where PageRank is not temporally biased and $r(p,\eta)/r(k^{in},\eta)$ becomes close to, albeit always strictly lower than, one.
When moving away from this diagonal,
PageRank score has temporal bias towards recent or old nodes (Fig. S6), 
its correlation with indegree (Fig. S7) and fitness (Fig. S8) decrease, and it reproduces
fitness substantially worse than indegree (red areas in Fig. \ref{fig:RM}). 
Qualitatively similar behavior is found for the RM with 
uniformly distributed fitness (Fig. S9), power-law decay of relevance and activity (Fig. S10),
accelerated growth rate ($m(t)\propto t$ instead of $m(t)=10$, Fig. S12).
The same is true when the ranking quality is measured by the precision metric $P_{100}(\cdot,\eta)$, 
(defined as the number of fitness top-$100$ nodes placed in the top $100$ of the
ranking produced by an algorithm), instead of the linear correlation coefficient (Fig. S11).
This shows that our findings are robust and do not require a specific model setting.

\begin{figure}[t]
\centering
\includegraphics[width=0.8\columnwidth, angle=0]{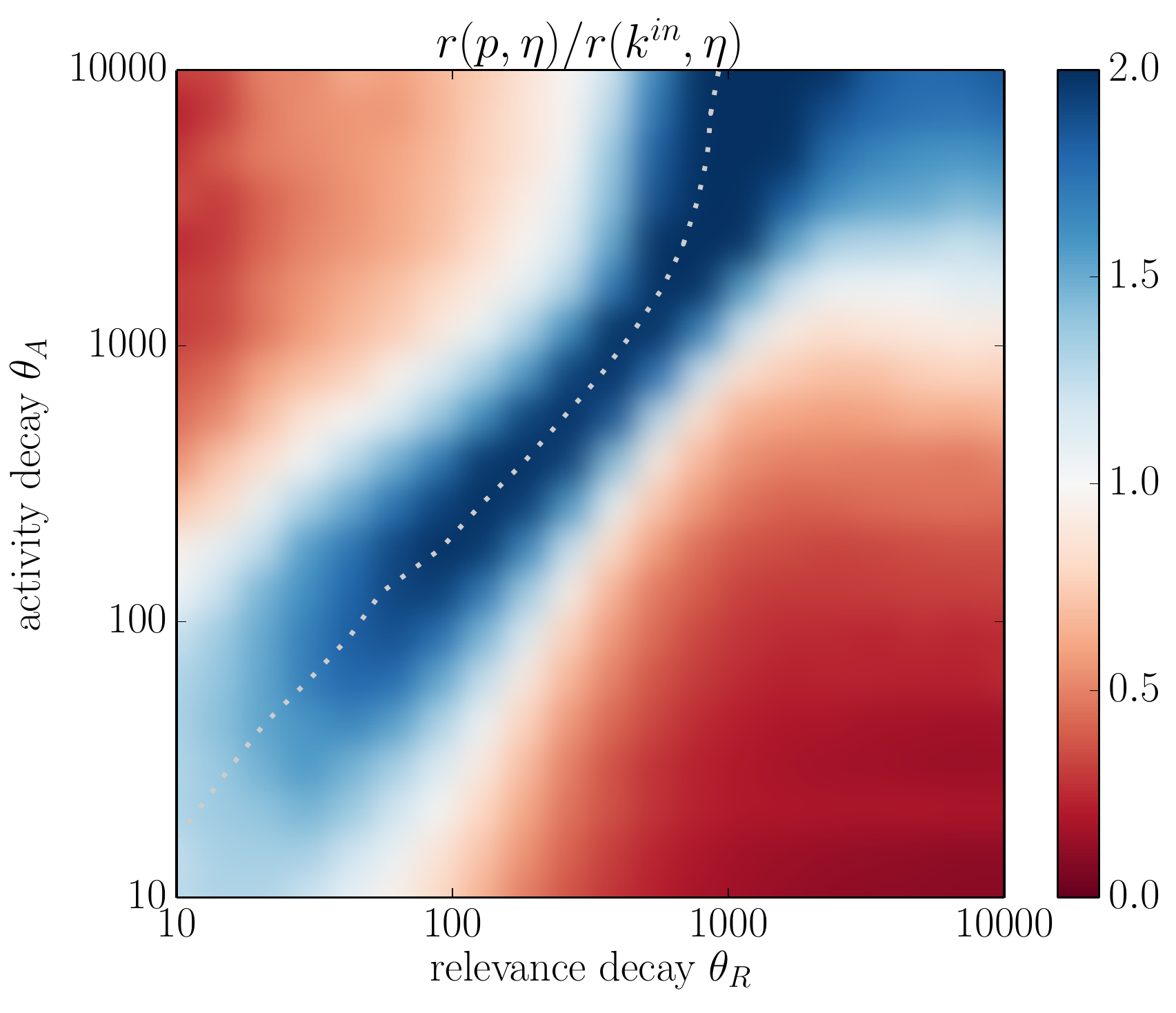}
\caption{\textbf{A comparison of PageRank and indegree correlation with fitness in the EFM data} ($N=10,000$, $H=250$, color online).
The heatmap shows the ratio $r(p,q)/r(k^{in},q)$. The white dotted line represents the contour where PageRank is not temporally biased
(see Fig. S6, right). \label{fig:EFM}}
\end{figure}

\subsection{An extended model based on fitness}

To demonstrate that PageRank's under-performance with respect to indegree is a general feature, we now proceed to a different model for artificial data
which is more compatible with PageRank's basic idea that a node is important if it is pointed by other important nodes.
In this model (hereafter Extended Fitness Model, EFM), high- and low-fitness nodes differ not only in their ability
to attract new incoming links, but also in their sensitivity to the fitness of the other nodes when choosing their outgoing connections.
High-fitness nodes are highly attractive to new incoming links as well as highly
sensitive to fitness of the others
when choosing their outgoing connections.
Low-fitness nodes are basically insensitive to fitness and choose their target nodes mostly by current popularity amended by aging.
High-fitness nodes are then more likely to be pointed by other high-fitness nodes than low-fitness nodes (see Fig. S5)
which agrees with the basic premise of PageRank: important nodes are pointed by other important nodes.
We therefore expect PageRank to outperform indegree in ranking the nodes by fitness.
The model assumes that the probability
$\Pi^{in}_{i;j}$ that a link created by node $j$ at time $t$ ends in node $i$ has the form
\begin{equation}
  \Pi^{in}_{i;j}(t)\sim (k_{i}^{in}(t)+1)^{1-\eta_{j}}\,\eta_{i}^{\eta_{j}}\,f_{R}(t-\tau_{i})
 \label{PAQD}
\end{equation}
where node fitness $\eta$ is now constrained to the range $[0,1]$ to prevent a negative exponent $1-\eta$ in the first term.
We stress that the probability $\Pi^{in}$ depends not only on the fitness $\eta_i$ of the target node,
but also on the fitness $\eta_j$ of the node $j$ that creates the outgoing link, which is a new element with respect
to the RM.
A similar model has been used to model user-item networks in \cite{medo2014inpreparation}.
We assume that a small number $H$ of nodes have high fitness ($\eta\in[10^{-5},1]$) and the remaining $N-H$ nodes 
have
low fitness ($\eta\in[0,10^{-5})$, see the Method section for details).

Fig. \ref{fig:EFM} shows the results obtained with the EFM. The correlation coefficient $r(p,k^{in})$ (Fig. S7, right) 
and the average age of top $1\%$ nodes (Fig. S6, right) have qualitatively the same behavior as for the RM
 which indicates that the 
behaviour of these quantities as a function of model's temporal parameters is universal and independent of the exact growth rule.
The model is favorable to PageRank and indeed, the algorithm now can significantly outperform indegree 
in terms of the correlation between fitness and node score when PageRank is not temporally biased (blue area in Fig. \ref{fig:EFM}).
Nevertheless, PageRank still underperforms indegree in two extensive regions of the parameter plane $(\theta_R,\theta_A)$.
As for the RM, these two regions correspond to the cases where activity and relevance decay timescales substantially differ.
These results are again confirmed by using power-law aging
instead of exponential (Fig. S10) and the precision metrics instead of the correlation coefficient (Fig. S11).
Note that we introduced here the EFM to show that PageRank's bias occurs also in a setting favorable to the algorithm;
while it seems plausible that some nodes are more sensitive to fitness than others when making connections, we leave real data validation of the EFM
for future research.

\begin{figure}[t]
\centering
\includegraphics[height=0.7\columnwidth, angle=270]{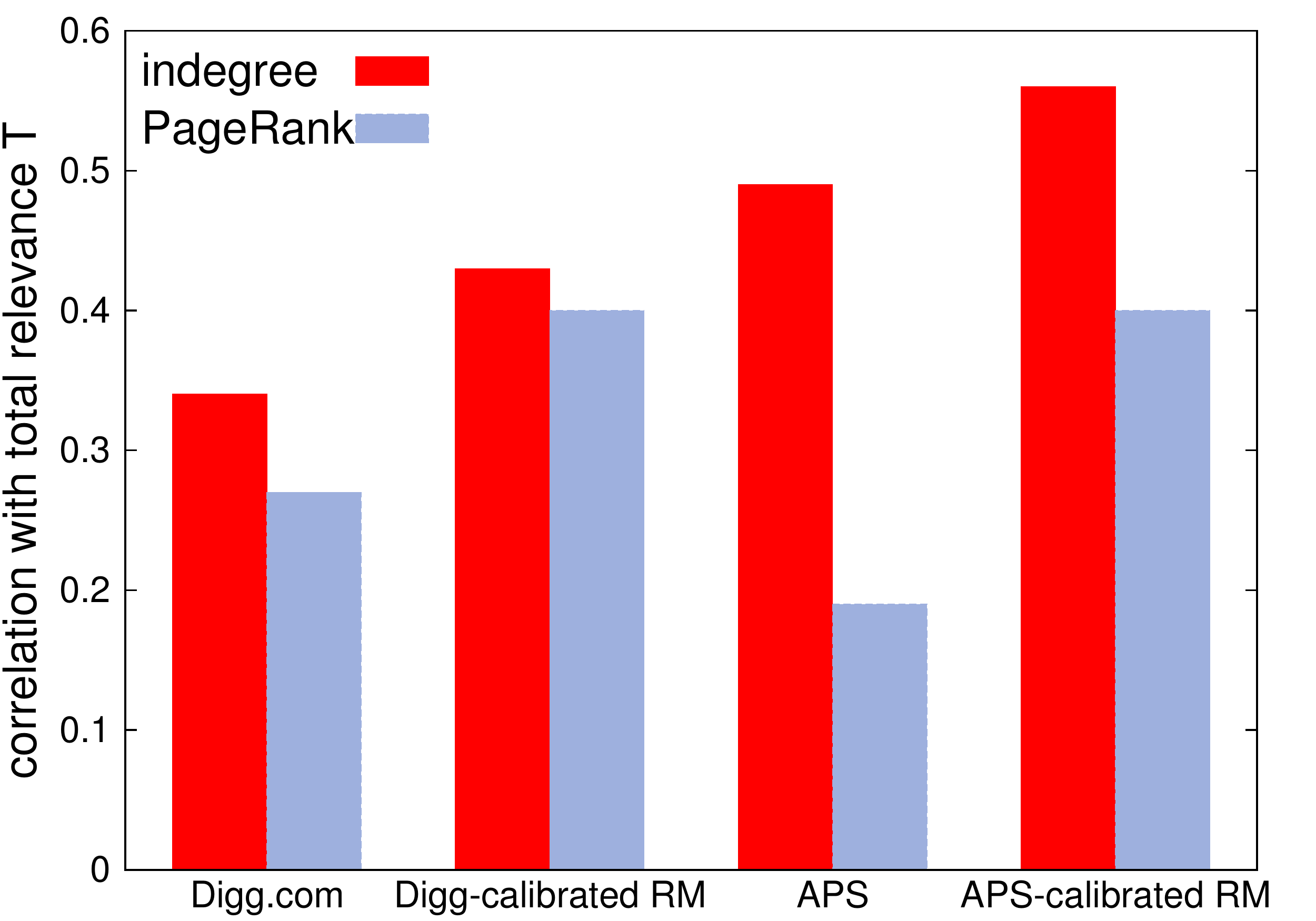}
\caption{\textbf{A comparison of PageRank and indegree correlation with total relevance in real data and in calibrated simulations with the RM.}
PageRank is outperformed by indegree in both datasets (and in the corresponding calibrated simulations). 
In the Digg.com social network, the fitted relevance and activity power-law decay exponents are not far from the parameter region where PageRank is maximally correlated with
indegree in numerical
simulations with the RM with power-law decay
(see Fig. S10),
and PageRank's and indegree's correlation with total relevance
are close to each other. By contrast, in the APS dataset activity decays immediately, 
whereas relevance decays progressively (see Fig. S2); as a
consequence, PageRank is strongly biased towards old nodes (see Fig. S3) and is outperformed 
by indegree by a factor $2.58$ [$r(p,T)=0.19$ whereas $r(k^{in},T)=0.49$].
We refer to the Supplementary Note S3 for details about the simulation calibration on real data and 
to the Supplementary Note S4 for details on the computation of empirical relevance in real and artificial data.
\label{fig:bars}}
\end{figure}

\subsection{Comparing indegree and PageRank: results in real networks}

Algorithm evaluation in real data is made difficult by several factors.
In general, it is impossible to objectively evaluate node importance in a system because it depends on many
intangible and subjective elements \cite{franceschet2011pagerank}.
To assess the performance of ranking algorithms on real data, we compare node 
score with \emph{total relevance} $T_{i}=\sum_{t}r_{i}(t)$ which is an estimate of node fitness 
(see Ref. \cite{medo2011temporal} and the Supplementary Note S4).
Results on real data and the corresponding calibrated simulations with the RM are reported in Fig. \ref{fig:bars}.
Our calibration procedure for simulations focuses on temporal decay of relevance and activity and is described in detail in the Supplementary Note S3;
more accurate calibration is possible
but goes beyond the scope of our work.
Uncertainty of these results estimated by sample-to-sample fluctuations and non-parametric bootstrap \cite{shalizi2010bootstrap} for model and 
real data, respectively,
is of the order of $10^{-3}$ which is negligible in comparison with the observed differences between PageRank and indegree
(see Supplementary Note S6).

In the Digg.com social network, the empirircal relevance and activity power-law decay exponents are not far from the parameter region where 
PageRank scores are maximally 
correlated with indegree in the
simulations with the RM with power-law decay
(see Fig. S10), which is in qualitative agreement with the observed high value of correlation between PageRank and indegree in the dataset
($r(k^{in},p)=0.88$); PageRank is outperformed by indegree in ranking nodes by their total relevance but the performances of the two metrics 
are relatively close to each other (see Fig. \ref{fig:bars}).

In citation data,
where the use of PageRank and other
algorithms inspired by PageRank has been much studied \cite{chen2007finding,walker2007ranking, maslov2008promise},
activity and relevance decays necessarily mismatch: relevance progressively decays with time \cite{medo2011temporal}, 
whereas activity decays immediately.
In the APS dataset we find that PageRank is significantly biased towards old nodes (Fig. S3):
this is because old papers can be pointed by papers of every age, while recent papers are pointed only by recent papers.
This is the opposite time bias than that depicted in Fig. \ref{fig:bias1}.
Moreover, we find that
PageRank and indegree are weakly correlated [$r(k^{in},p)=0.52$], and indegree is remarkably better correlated with total relevance than PageRank
(see Fig. \ref{fig:bars}). These findings are consistent with the outcomes of a calibrated numerical simulation
with the RM (see Fig. \ref{fig:bars}), where all outgoing links of a node are created when the node enters the system and the
outdegree distribution is exponential
as in the APS dataset (Fig. S4), see the Supplementary Note S3 for details about simulations calibrated on real data).
Note that the age distribution of top nodes in indegree and PageRank ranking 
in the
APS real network and the artificial network generated by the corresponding calibrated simulation  
have the same qualitative shape (Fig. S3).
This confirms that our simulation calibration on real data, which is based only on the temporal patterns of the system,
qualitatively captures the temporal bias of PageRank.

We conclude this paragraph with a consideration on
total relevance $T$. Motivated by the high correlation $r(T,\eta)$ between node total relevance and fitness found in the calibrated simulations
(see Supplementary Note S4),
in this work we use total relevance $T$ as a proxy for node fitness in the real data.
In the RM, we find that node total relevance outperforms indegree and PageRank in ranking nodes by fitness for
a broad range of model parameters (Fig. \ref{SIfig:tot_rel_rm}). 
By contrast, the parameter region where total relevance outperforms indegree and PageRank is smaller 
in data produced with the EFM (Fig. \ref{SIfig:tot_rel_efm}). We leave for future research detailed investigation of how the 
performance of total relevance in ranking nodes by fitness depends on the assumptions and parameters of the underlying model.
These findings might also motivate future study of the ranking of nodes by their total relevance in real data that are well-described by the RM.

\section{Discussion}

To summarize, our numerical simulations indicate that the mismatch between the timescales of relevance and activity 
decay makes PageRank scores biased towards recent nodes
(when the decay of relevance is faster)
or old nodes (when the decay of activity is faster). 
This temporal bias reduces PageRank's capability to rank nodes by fitness and causes it to underperform in comparison with the elementary ranking of nodes by 
indegree in the RM which is to our best knowledge the most accurate model for describing growing
information networks \cite{wang2013quantifying, medo2014statistical}. 
Our findings are
robust with respect to changes in the functional form of the time-decay function, in
the distribution of fitness among the nodes, and in the metric used to evaluate the ability of an algorithm
to rank nodes by their fitness. 
We also studied a model (the EFM) that provides a favorable setting for PageRank performance;
PageRank can outperform indegree on the data produced by this model, but fails again when the two timescales mismatch.          
Moreover, 
we find indications of the influence of temporal patterns on PageRank's performance also in real data. In citation data,
PageRank is excessively biased towards old nodes and, as a consequence, is clearly outperformed by indegree in ranking nodes by their total relevance which is
an estimate of node fitness (see Ref. \cite{medo2011temporal} and the Supplementary Note S4).     
By contrast, 
indegree and PageRank perform similarly in social network data where there is not a sharp mismatch between activity and relevance timescales.
The results of real data analysis are in agreement with our model-based finding that PageRank can only 
perform well if the two system's timescales (of activity and relevance decay, respectively) are of similar magnitude.

The methods developed and used in this article are general and can be applied to any growing directed
network where nodes compete for incoming links
and where preferential attachment and temporal effects influence the linking patterns, which includes a wide class of real networks.
To diagnose whether a growing directed network
is or not suitable for the application of PageRank, one can fit the empirical relevance decay and activity decay timescales 
on the data and run a corresponding calibrated simulation
which reveals whether PageRank is or not able to rank nodes according to their fitness. 
We have not attempted to study how our findings are affected by further real-world
phenomena, such as link deletion \cite{kong2008experience}, popularity \cite{ratkiewicz2010characterizing} 
and activity bursts \cite{barabasi2005origin}, among others.
Link time stamps are crucial for our analysis; in all the datasets where they are not reported,
we cannot compute neither node relevance $r(t)$ nor node activity $a(t)$
which exclude these systems from the range of applicability of our analysis.
We also stress that the framework introduced in this work is not 
applicable to undirected networks, such as collaboration networks, scholar co-citation networks and road networks,
among others.
In undirected networks indeed 
there is no distinction between incoming and outgoing links
and, as a consequence, relevance and activity cannot be defined as two separate node properties.
The model-based evaluation of PageRank's performance in networks without time information and undirected networks
is certainly an interesting and largely unexplored problem but
goes beyond the scope of this work.

The shortcoming of PageRank due to temporal effects 
is particularly worrying for applications of the algorithm to scientific citation 
data \cite{bollen2006journal,maslov2008promise,ma2008bringing}.
While PageRank can find old valuable papers underestimated by indegree \cite{chen2007finding},
the algorithm is biased towards old nodes
and as a consequence is outperformed by indegree in ranking papers by importance, which strongly discourage the use of the algorithm to rank scientific papers.
In this context, the need for including temporal effects in the algorithm has already been stressed in \cite{walker2007ranking, maslov2008promise};
the model-based approach introduced in this article leads us to the same conclusion.
How to best include the temporal dimension in ranking scientific publications remains an open issue. One could consider a self-consistent algorithm
that takes time into account, such as CiteRank \cite{walker2007ranking}, or resort to fitness estimates, such as total relevance or maximum likelihood 
estimates \cite{medo2014statistical}; 
the model-based approach introduced in this article provides a simple yet effective
method -- the comparison of node scores with intrinsic fitness in calibrated simulations
-- which could be used to establish which algorithm is more suitable for a given system.
Our findings also bring new insights into the study of the relation between node indegree and PageRank. Previous
studies \cite{fortunato2006topical,fortunato2008approximating} established a linear relationship between node degree and the average PageRank score
for uncorrelated networks, and 
considered any deviations from this behavior as fluctuations. We find that for a broad range of network parameters, 
much of these apparent fluctuations are in fact trends caused by the interplay between the network's temporal features and the PageRank algorithm.

Our model-based evaluation of ranking algorithm is applicable also to the WWW.
There is general agreement in recognizing the importance of PageRank in the success of Google's search engine \cite{franceschet2011pagerank,langville2011google},
yet it remains unclear which properties make the Web a suitable network where to apply the algorithm.
While Ref. \cite{ghoshal2011ranking} emphasizes the role of the scale-free topology of the Web on PageRank's success,
our
findings stress the importance of temporal patterns in determining the success or failure
of PageRank. 
Further data analysis on Web data could reveal whether relevance and activity decay timescales are of similar magnitude
in the WWW which would imply maximal correlation between PageRank score and node fitness, and thus provide a further 
explanation of PageRank's success in this system.

In conclusion, PageRank, despite its popularity and robustness, can fail and thus it should not be used without carefully considering the 
temporal properties
of the system to which it is to be applied.
The connection between PageRank's failure and the temporal features of the analyzed networks indicates that the main reason for the reported failure is the 
static nature of the algorithm.
We believe that a well-grounded ranking algorithm should be built on the temporal patterns  
of the system where it is intended to be applied and the dependence of its performance on system features should be exhaustively studied in model data
where system's structural and temporal properties can be modified simply by changing model parameters.
We believe that 
the model-based theoretical evaluation of ranking algorithms developed in this work 
will open the door to systematic performance evaluation of algorithms in evolving systems, 
deepen our understanding of their limitations, and lead to the introduction of new improved algorithms.

\section{Methods}

 \paragraph*{Digg.com dataset.} Digg.com had been an online social news aggregator from December 2004 to July 2012. 
Digg.com users were allowed to submit and vote (``digg'') stories.
Interaction between users took place through comments
and messages (see \cite{doerr2012friends} for a detailed description of the website).
We studied the social network of users
where nodes represent the users and a link from node $i$ to node $j$ means that user $i$ is a follower of user $j$.
The complete dataset in our possession covers the period from $06/08/2005$ to $08/07/2009$.
We analyzed a $3$-years subset running from $01/01/2006$ to $31/12/2008$.
The subset consists of $N=190,553$ nodes and $L=1,552,905$ links.

 \paragraph*{APS dataset.} The APS (American Physical Society) dataset in our possession spans from year 1893 until 2009 and 
 contains $N=450,056$ nodes (papers) and $L=4,690,967$ directed links (citations) between them.
This dataset has been used in \cite{medo2011temporal} to validate the hypothesis of heterogeneous and decaying relevance. 

\paragraph*{PageRank.} In a directed monopartite network composed of $N$ nodes where no dangling nodes (nodes with zero outdegree) exist,
the vector of PageRank scores $\{p_{i}\}$ can be found as the stationary solution of the following set of recursive linear equations
\begin{equation}
 p_{i}^{(t+1)}=c\,\sum_{j}A_{ji}\,\frac{p_{j}^{(t)}}{k^{out}_{j}}+\frac{1-c}{N},
\label{pr}
\end{equation}
where $A$ is the network's adjacency matrix ($A_{ji}$ is one if node $j$ points to node $i$ and zero otherwise), $k^{out}_{j}$ is the outdegree of node $j$,
$c$ is the teleportation parameter, and $t$ is the iteration number \cite{brin1998anatomy} .
Eq. \eqref{pr} represents the master equation of a diffusion process on the network, which converges to a unique stationary state 
independently of the initial condition \cite{berkhin2005survey}.
The PageRank score
$p_{i}$ of node $i$ can be interpreted as the average fraction
of time spent on node $i$ by a random walker who
with probability $c$ follows the network's links and with probability $1-c$ teleports to a random node.
We set $c=0.85$ which is the usual choice in practice \cite{berkhin2005survey}.
Iterations are stopped when the modulus distance between the vectors of scores at two consecutive iterations becomes smaller 
than $\epsilon=10^{-8}$ \cite{berkhin2005survey}.

\paragraph*{Simulation details.} We use the artificial models (RM and EFM) to build monopartite directed networks composed of $N=10,000$ nodes. 
We start from a configuration with two nodes, node $0$ and node $1$,
and a link from node $1$ to node $0$. At each simulation step $t$, we add a new node $t$ to the system and connect it to an already existing node.
The target node is chosen according to the attachment rule \eqref{PARD} (RM) or \eqref{PAQD} (EFM).
If $t>10$, we also sequentially add $m=10$ links between the existing nodes. Their initial nodes are chosen
according to the activity rule \eqref{activity}; the target nodes follow again Eqs. \eqref{PARD} or \eqref{PAQD}, respectively.
The creation of multiple links between a pair of nodes and self-loops are prohibited.
Unless stated otherwise, results are averages over $6$ realizations of the model. Error bars
In Fig. \ref{fig:bias2} represent the standard error of the mean which is generally small.
The same is true for Figures \ref{fig:RM} and \ref{fig:EFM} where only the average values are displayed.

\paragraph*{Fitness and activity distributions in the RM.} When relevance decay is sufficiently fast to allow the
normalization factor $\Omega(t)$ of $\Pi^{in}$ to converge within the simulation time scale \cite{medo2011temporal},
the average final indegree of a node in the RM depends exponentially on node fitness.
Consequently, different fitness distributions yield different indegree distributions \cite{medo2011temporal,bianconi2001competition}.
We use both exponential and uniform fitness distribution in our simulations; results for the latter are shown in Fig. S9.
The outdegree distribution is only determined by the activity distribution $\rho(A)$ (see Supplementary Note S7 for basic analytical results).
In our simulations we use $\rho(A)=2\, A^{-3}$ for $A\in[1,\infty)$ everywhere except for the calibrated APS data simulation where
all outgoing links of a node are created when the node enters the system and we use $\rho(k)=8.33\,\exp{(-0.12\,k)}$ for $k\in[0,\infty)$
as the outdegree distribution, as 
found in the APS data (see Fig. S4).

\paragraph*{Fitness and activity distribution in the EFM.} We choose here a fitness distribution that aims to 
emphasize the difference between the linking pattern of high- and low- fitness nodes without trying to reproduce structural features of real data.
The set of fitness values consists of
$N-H$ equidistant values within the interval $[0,\eta_{th})$ (low-fitness nodes) and $H$ equidistant values from the range $[\eta_{th},1]$
(high-fitness nodes).
These values are then bijectively assigned to the network's $N$ nodes at random.
We set a small value of the threshold $\eta_{th}=10^{-5}$ which implies that the low-fitness nodes are essentially insensitive to node fitness,
while the high-fitness nodes range from little fitness-sensitive nodes to nodes almost unaffected by popularity and mainly driven by fitness
(when $q_{j}=1$, we have $\Pi^{in}_{i}\sim q_{i}\,f_{R}(t-\tau_{i})$). 
We run simulations with $H=250=N/40$ for Fig. \ref{fig:EFM}.
this value is small in order to amplify the advantage of high-fitness nodes in connecting to other high-fitness nodes (see Fig. S5).
As in the RM, we use $\rho(A)=2\, A^{-3}$ for $A\in[1,\infty)$ to generate the node activity values.

\bibliographystyle{naturemag}

\paragraph*{Acknowledgements.}
This work was supported by the EU FET-Open Grant No. 611272 (project Growthcom). 
We acknowledge useful discussions with An Zeng and Hao Liao.
The authors declare that they have no competing financial interests.
 Correspondence and requests for materials should be addressed to M. S. Mariani~(email: manuel.mariani@unifr.ch).

\clearpage

\setcounter{figure}{0}
\makeatletter 
\renewcommand{\thefigure}{S\@arabic\c@figure}
\renewcommand{\theequation}{S\@arabic\c@equation}
\makeatother

\section*{Supplementary notes}

\paragraph*{S1: Analysis of empirical relevance and activity in the Digg.com dataset.}
\label{sec:Digg_fit}
As explained in the main text, the empirical relevance $r_{i}(t)$ of node $i$ at time $t$ is defined as
\begin{equation}
\label{SIempiricalrelevance}
r_{i}(t)=\frac{n_{i}(t)}{n^{PA}_{i}(t)}
\end{equation}
where $n_{i}(t)=\Delta k^{in}_{i}(t,\Delta t)/L(t,\Delta t)$ is the ratio between the number $\Delta k^{in}_{i}(t,\Delta t)$ of incoming links received 
by node $i$ in a suitably chosen time window $[t,t+\Delta t)$ and the total number $L(t,\Delta t)$ of links created within the same time window,
whereas $n^{PA}_{i}(t)=k^{in}_{i}(t)/\sum_{j}k^{in}_{j}(t)$ is the expected value of $n_{i}(t)$ according to preferential attachment alone~\cite{medo2011temporal}.
The activity $a_{i}(t)$ of node $i$ at time $t$ has been defined in \cite{perra2012activity} as
\begin{equation}
\label{SIempiricalactivity}
a_{i}(t)=\frac{\Delta k^{out}_{i}(t,\Delta t)}{L(t,\Delta t)}
\end{equation}
where $\Delta k^{out}_{i}(t,\Delta t)$ is the number of outgoing links created by node $i$ in the time window $[t,t+\Delta t)$. We use $\Delta t=1\,\text{week}$ to compute relevance and activity in the Digg.com dataset.

\begin{figure*}[h]
\centering
\includegraphics[scale=0.27, angle=270]{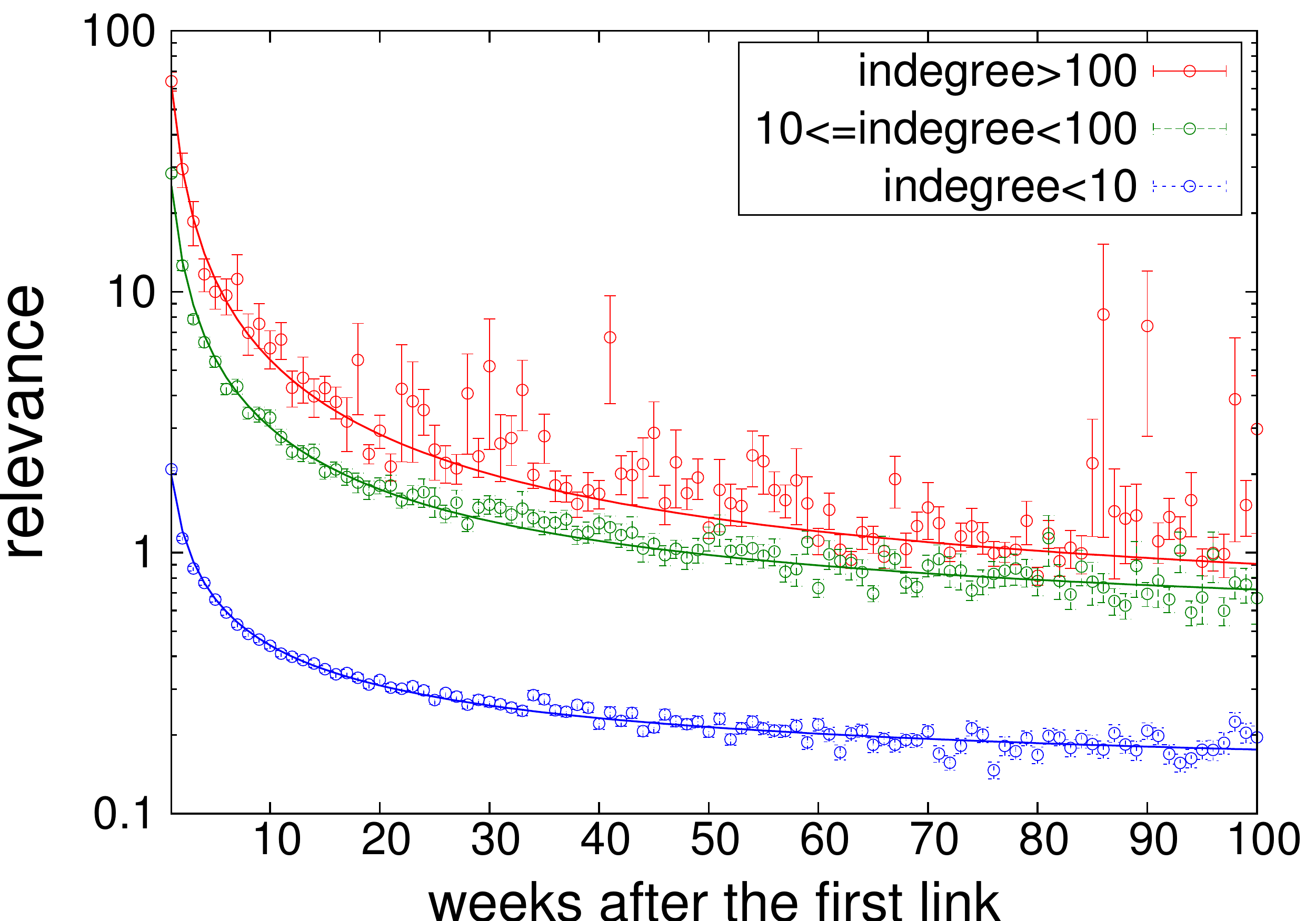}\qquad\includegraphics[scale=0.27, angle=270]{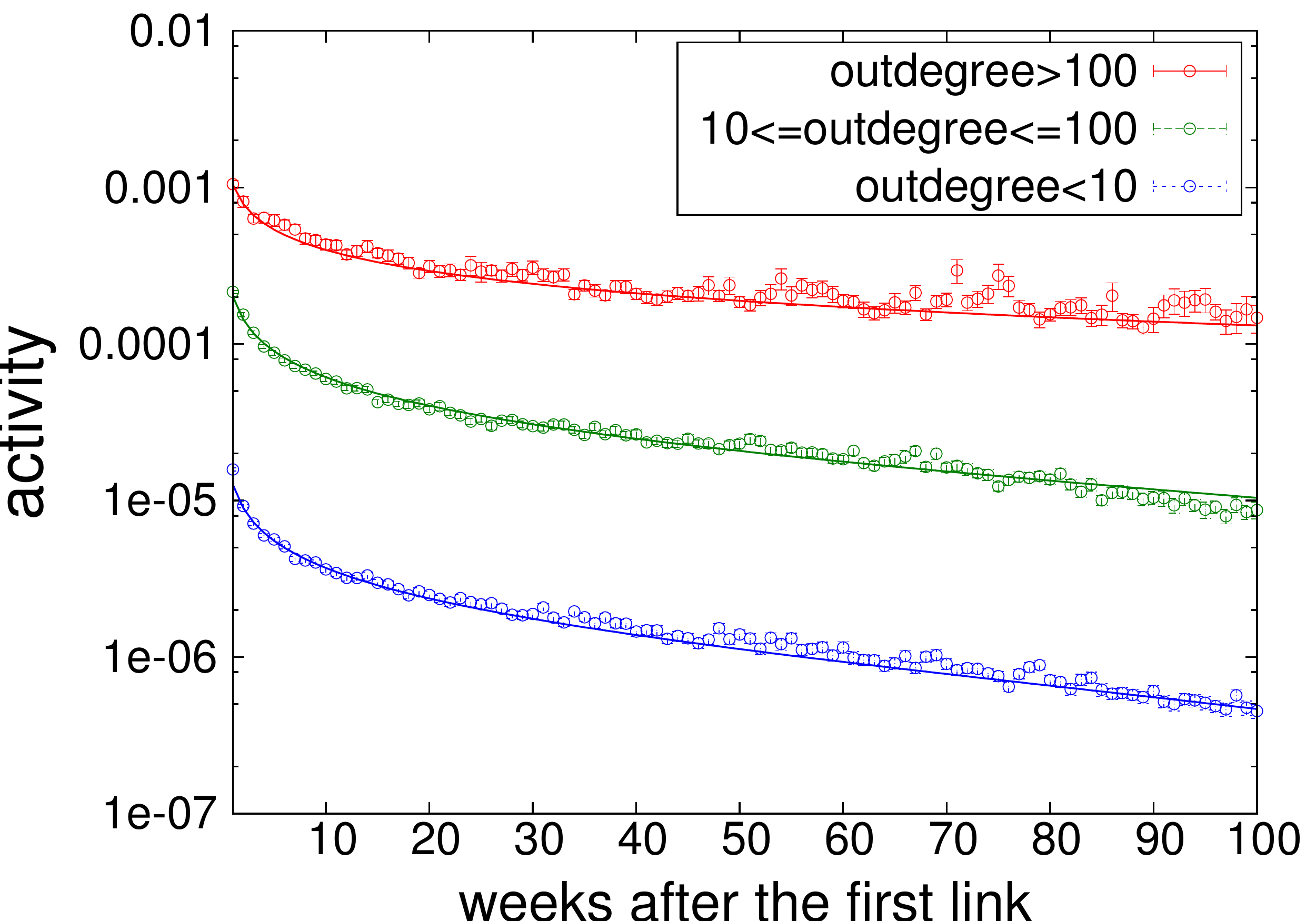}
\caption[Temporal decay of the average relevance $r(t)$ and activity $a(t)$ in Digg.com social network]{\textbf{Temporal decay of the average
relevance $r(t)$ (left panel) and activity $a(t)$ (right panel) in Digg.com social network} (2006-2008, $\Delta t=1\,\text{week}$, color online). Symbols represent the average relevance and activity of nodes belonging to the same age group, error bars represent the errors of the mean, lines represent the fits described in Section~\ref{sec:Digg_fit}.}
\label{SIfig:digg_decay}
\end{figure*}

Figure \ref{SIfig:digg_decay} shows the average temporal decay of relevance and activity in the Digg.com dataset (see MM section of the main text for a description of the dataset). We fit both decay profiles with a power-law function $f(t)= C\,t^{-\alpha}+D$ with three parameters ($C$, $\alpha$, $D$) using the least-squares method.
While other and perhaps more accurate patameter estimation procedures exist, the present results are sufficient for our analysis. We only consider the first $100$ weeks after the first link received/created by the respective node. Parameter estimates are summarized in Table~\ref{tab:fitting_digg} separately for nodes of high, medium and low in-degree and out-degree, respectively. One may note here that activity decays slower (with a lower exponent) than relevance.

\begin{table}
\centering
\begin{tabular}{lrrr}
\multicolumn{4}{c}{\it Relevance decay}\\
\hline
         Node group & $C$ & $\alpha$ & $D$\\
\hline
       $k^{in}>100$ & $61.31$ & $1.09$ & $0.50$\\
$k^{in}\in[10,100]$ & $25.00$ & $0.99$ & $0.46$\\
        $k^{in}<10$ &  $1.85$ & $0.76$ & $0.12$\\
\hline
\end{tabular}
\qquad
\begin{tabular}{lrrr}
\multicolumn{4}{c}{\it Activity decay}\\
\hline
          Node group & $C$ & $\alpha$ & $D$\\
\hline
       $k^{out}>100$ & $1.1\cdot 10^{-3}$ & $0.38$ & $-6.0\cdot 10^{-5}$\\
$k^{out}\in[10,100]$ & $2.2\cdot 10^{-4}$ & $0.44$ & $-1.86\cdot 10^{-5}$\\
        $k^{out}<10$ & $1.4\cdot 10^{-5}$ & $0.44$ & $-1.38\cdot 10^{-6}$\\
\hline
  \end{tabular}
\caption{Parameter estimation results for the average relevance and activity decay in the Digg.com data.}
\label{tab:fitting_digg}
\end{table}

\begin{figure*}[h]
\centering
\includegraphics[scale=0.27, angle=270]{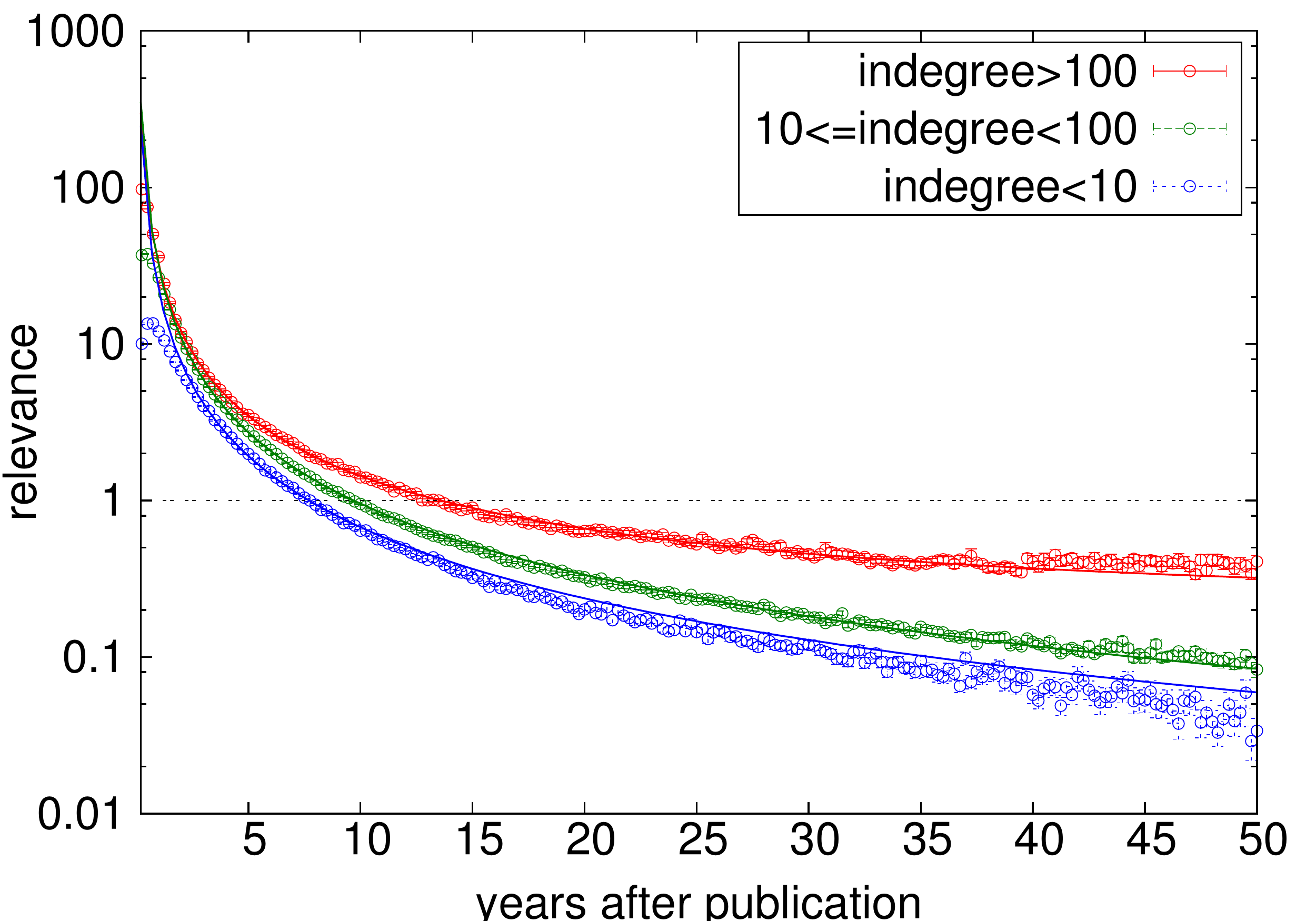}
\caption[Temporal decay of the average relevance $r(t)$ in the APS dataset]{\textbf{Temporal decay of the average relevance $r(t)$ in the APS dataset} (1893-2009, $\Delta t=91\,\text{days}$, color online). Symbols represent the average relevance of nodes belonging to the same age group, error bars represent the error of the mean, lines represent the fits described in Section~\ref{sec:APS_fit}. The initial non-monotonous part of the relavance profile is ignored by the fitting procedure and consequently the fitted curves do not match the points corresponding to the first few years after publication.}
\label{SIfig:aps_decay}
\end{figure*}

\paragraph*{S2: Analysis of empirical relevance in the APS dataset.}
\label{sec:APS_fit}
We use $\Delta t=91\,\text{days}$ to calculate empirical relevance of nodes in the APS dataset (see MM section of the main text for a description of the dataset). Figure~\ref{SIfig:aps_decay} shows the average relevance decay in the APS dataset; it is analogous to Figure~1 in~\cite{medo2011temporal}. Similarly as for the Digg.com dataset, we fit the results with the power law dependence $f(t)=C\,t^{-\alpha}+D$. To avoid the non-monotonous initial behavior of relevance (which is due to, for example, the time needed to carry out and publish research building on a given paper), we ignore the first $5$ years ($10$ years for low indegree nodes) after publication. The estimation results are reported in Table~\ref{tab:fitting_aps}.

\begin{table}
\centering
\begin{tabular}{lrrr}
\multicolumn{4}{c}{\it Relevance decay}\\
\hline
      Node group    & $C$ & $\alpha$ & $D$\\
\hline
       $k^{in}>100$ & $217.5$ & $1.40$ & $0.19$\\ 
$k^{in}\in[10,100]$ & $250.8$ & $1.51$ & $0$\\
        $k^{in}<10$ & $177.2$ & $1.51$ & $0$\\
\hline
\end{tabular}
\caption{Parameter estimation results for the average relevance decay in the APS data.}
\label{tab:fitting_aps}
\end{table}

\begin{figure*}[h]
\centering
\includegraphics[scale=0.27, angle=270]{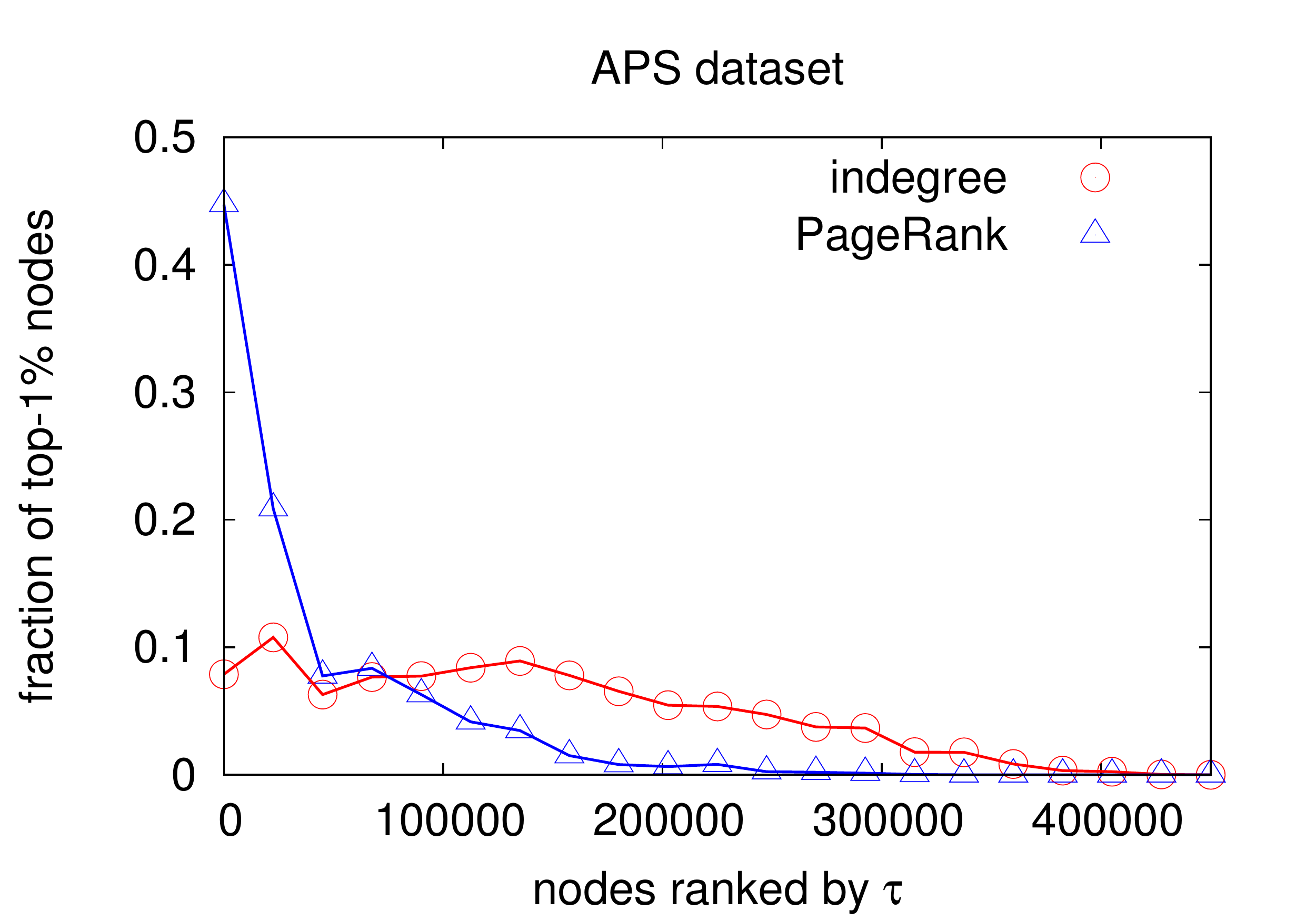}\qquad\includegraphics[scale=0.27, angle=270]{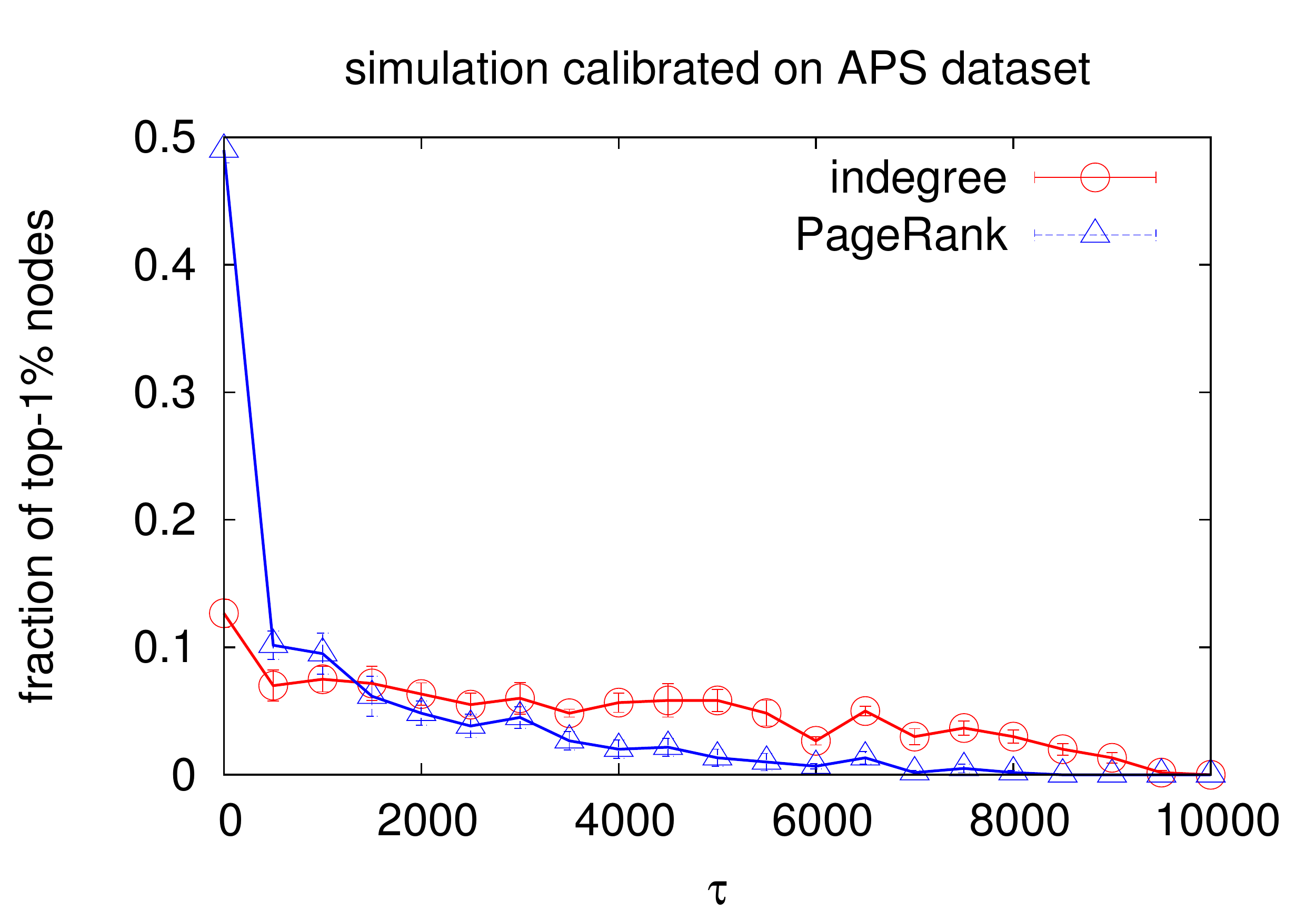}
\caption[Age distribution of the top $1\%$ nodes in the ranking (APS data and the corresponding calibrated simulation)]{\textbf{Age distribution of the top $1\%$ nodes in the ranking, in the APS data (left panel) and in the corresponding calibrated simulation (right panel)} (color online). On the $x$-axis, nodes are ordered by age (oldest on the left, youngest on the right). We see that PageRank is more biased towards old nodes than indegree. Results obtained on the calibrated model (see Table~\ref{tab:fitting_aps} for model parameters) agree well with those obtained on the real data.}
\label{SIfig:age}
\end{figure*}

\paragraph*{S3: Simulations calibrated on real data.}
\label{sec:calibration}
When calibrating the numerical simulations on the Digg.com and APS datasets, we focus only on the datasets' temporal patterns that constitute the main motivation of our study and are the principal reason for the reported failure of PageRank. While more accurate calibration of models to the real data is possible, we do not find it necessary because our calibrated simulations capture some basic temporal patterns of indegree and PageRank scores (see Figure~\ref{SIfig:age}).

\begin{figure*}[h]
\centering
\includegraphics[scale=0.27, angle=270]{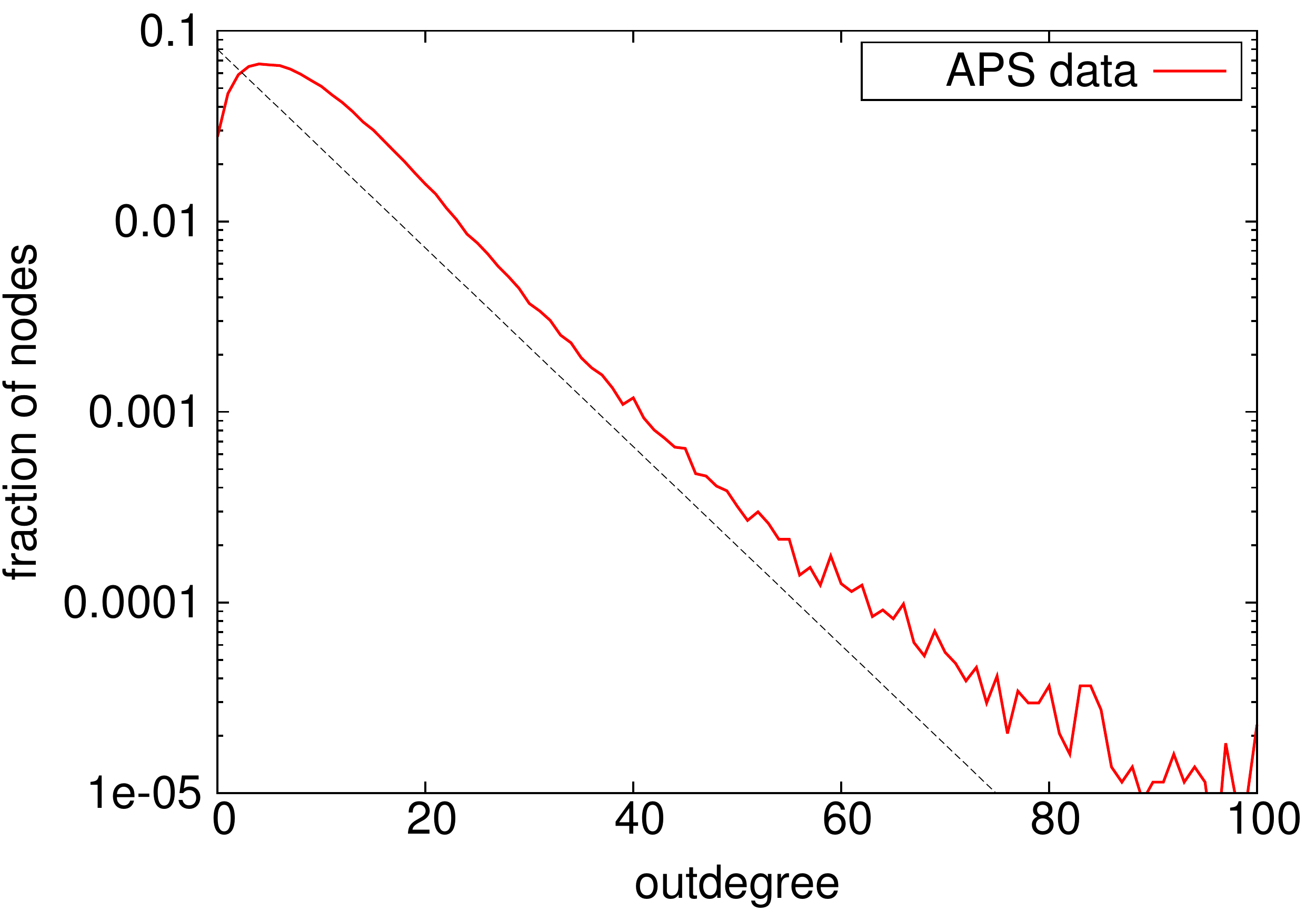}
\caption[Outdegree distribution in the APS dataset]{\textbf{Outdegree distribution in the APS dataset.} The distribution is narrow and is well approximated by $\rho(k^{out})\sim \exp(-0.12\,k^{out})$ for $k^{out}>5$ (black line). We use this distribution to generate the number of outgoing links created by a new node in the simulations calibrated on the APS dataset.}
\label{SIfig:outdegree}
\end{figure*}

The artificial dataset calibrated on the Digg.com data is grown using the relevance model (RM) with $\rho(\eta)=\exp(-\eta)$ and power-law decay of relevance ($\alpha_R=1$) and activity ($\alpha_A=0.4$); the other simulations details are the same described in the MM section of the main text. The artificial dataset calibrated on the APS data is grown using the RM with $\rho(\eta)=\exp(-\eta)$ and power law decay of relevance ($\alpha_R=1.4$); all outgoing links of a node are created when the node enters the system and the outdegree distribution is $\rho(k)=8.33\,\exp(-0.12\,k)$ as in the APS data (see Figure~\ref{SIfig:outdegree}).

\paragraph*{S4: Measuring empirical relevance in real and artificial data.}
\label{sec:relevance}
Since fitness values are not known in real data, we use the total relevance defined by Eq. \eqref{empiricalrelevance} as an estimator of node fitness. As shown for the RM in~\cite{medo2011temporal}, total relevance and fitness are closely connected and both provide information about the perceived importance of a node. However, the direct use of Eq. \eqref{empiricalrelevance} poses a problem in artificial data. In our numerical simulations, a constant number of link are added to the system at each time step. The factor $\sum_{j}k_{j}^{in}(t)$ on the right side of Eq. \eqref{empiricalrelevance} consequently grows linearly with simulation time and, as a result, the total relevance computed with Eq. \eqref{empiricalrelevance} is biased towards recent nodes. This issue does not occur in real data where both $L(t,\Delta t)$ and $\sum_{j}k_{j}^{in}$ grow with time. To avoid this bias, we omit the factor $\sum_{j}k_{j}^{in}$ when computing relevance in real data and use the following definition
\begin{equation}
\label{modifiedempiricalrelevance}
\tilde{r}_{i}(t; \Delta t)=\frac{n_{i}(t; \Delta t)}{k_{i}^{in}(t; \Delta t)}.
\end{equation}
The corresponding definition of total relevance for model data is $T_{i}(t; \Delta t):=\sum_{t}\tilde{r}_{i}(t; \Delta t)$; we use $\Delta t = 20$. This quantity is used in Figure~\ref{fig:bars} in the main text to compare the rankings by indegree and PageRank on calibrated artificial data. In these simulations, we find $r(T,\eta)=0.71$ and $r(T,\eta)=0.65$ for the RM calibrated on the Digg.com and APS dataset, respectively. High values of the correlation between $T$ and $\eta$ confirm that total relevance is a suitable estimator of a node's intrinsic fitness.

\paragraph*{S5: Production of Figure \ref{fig:bias1} in the main text.}
\label{sec:fig}
To produce Figure \ref{fig:bias1}, we start from a network with two nodes: node $0$ and node $1$, and one link between them (from $1$ to $0$). The final network consists of $N=40$ nodes and is grown up according to the RM. Node fitness is drawn from the exponentia distribution $\rho(\eta)=\exp(-\eta)$. Node relevance decays exponentially with $f_{R}(t)= (0.4)^{t}$. At each simulation step, a new node is added to the system and connected to an existing node according to Eq.~\ref{PARD} in the main text. Consequently, one new link is created among the existing nodes (so-called internal link). While the target node is again chosen according to Eq.~\ref{PARD}, the starting node is chosen at random from the existing nodes which corresponds to constant node activity.

\paragraph*{S6: Assessing the uncertainty of results.}
\label{sec:uncertainty}
Non-parametric bootstrap is a statistical method to estimate the error on quantities measured in real data \cite{shalizi2010bootstrap}. To estimate the errors of the correlation coefficients $r(k^{in},T)$ and $r(p,T)$ for a real dataset, we create new datasets by resampling with repetition from the given dataset. Since a resampled dataset can in principle contain multiple links between a pair of nodes, we compute PageRank on the resampled data using the generalized formula
\begin{equation}
\label{pr_iter}
p_{i}^{(t+1)}=c\,\sum_{j}M_{ji}\,\frac{p_{j}^{(t)}}{k^{out}_{j}}+\frac{1-c}{N},
\end{equation}
where $M_{ji}$ is the number of directed links from $j$ to $i$ (and correspondingly $k^{out}_{j}=\sum_{i}M_{ji}$). The correlation values of interest can be computed for resampled dataset. The standard deviation of these results over datasets then characterizes the uncertainty of the original correlation values. For the Digg.com data, we obtain $r(k^{in}, T)=0.330\pm0.003$ and $r(p,T)=0.224\pm0.003$ which means that the uncertainty is small and insignificant in comparison with the absolute differences between the correlation values. Results for the APS data lead to the same conclusion.

For the results obtained with calibrated simulations, we estimate their uncertainty by analyzing several model realizations and evaluating the standard error of the mean for a quantity of interest.
Using $50$ model realisations, we find $r(k^{in},T)=0.444\pm0.002$ and $r(p,T)=0.411\pm0.002$ which has the same implications as before: the results' uncertainties are substantially smaller than the absolute difference and thus unsignificant. Results for the APS data lead to the same conclusion.

\paragraph*{S7: Relation between outdegree and activity.} 
\label{sec:outdegree}
Eq.~\eqref{activity}, which governs the creation of outgoing links, does not contain the preferential attachment mechanism and as a consequence the final outdegree is determined only by node activity $A_{i}$. When activity decay $f_{A}$ is sufficiently fast to allow the normalisation factor of $\Pi^{out}$ to converge, the asymptotic solution in the continuum approximation \cite{newman2010networks} reads
\begin{equation}
\overline{k^{out}_{i}(t)}=1+m\,A_{i}\,\frac{\int_{\tau_{i}}^{t}dt'\,f_{A}(t'-\tau_{i})}{\Omega_{\infty}^{out}},
\label{activity1}
\end{equation}
where $\tau_{i}$ is the time at which node $i$ has entered the system and $\Omega_{\infty}^{out}=\lim_{t\to\infty}\sum_{i}A_{i}\,f_{A}(t-\tau_{i})<\infty$.
When activity decay is absent, this has an asymptotic solution
\begin{equation}
\overline{k^{out}_{i}(t)}=1+m\,\frac{A_{i}}{\overline{a}}\,\log\biggl(\frac{t}{\tau_{i}}\biggr)
\label{activity2}
\end{equation}
where $\overline{a}$ is the average node activity. The outdegree distribution is consequently determined mainly by the activity distribution $\rho(A)$.

\clearpage

\onecolumngrid

\section*{Supplementary figures: Analysis of real data}

\clearpage

\onecolumngrid

\section*{Supplementary figures: Numerical simulations}

For the following figures, unless stated otherwise, the models' settings are:
\begin{itemize}
 \item RM: $N=10,000$, $\rho(\eta)=\exp{(-\eta)}$, $f_{R}(t)=\exp{(-t/\theta_R)}$, $f_{A}(t)=\exp{(-t/\theta_A)}$.
 \item EFM: $N=10,000$, $H=250$, $f_{R}(t)=\exp{(-t/\theta_R)}$, $f_{A}(t)=\exp{(-t/\theta_A)}$.
\end{itemize}

\begin{figure*}[h]
\centering
\includegraphics[scale=0.75]{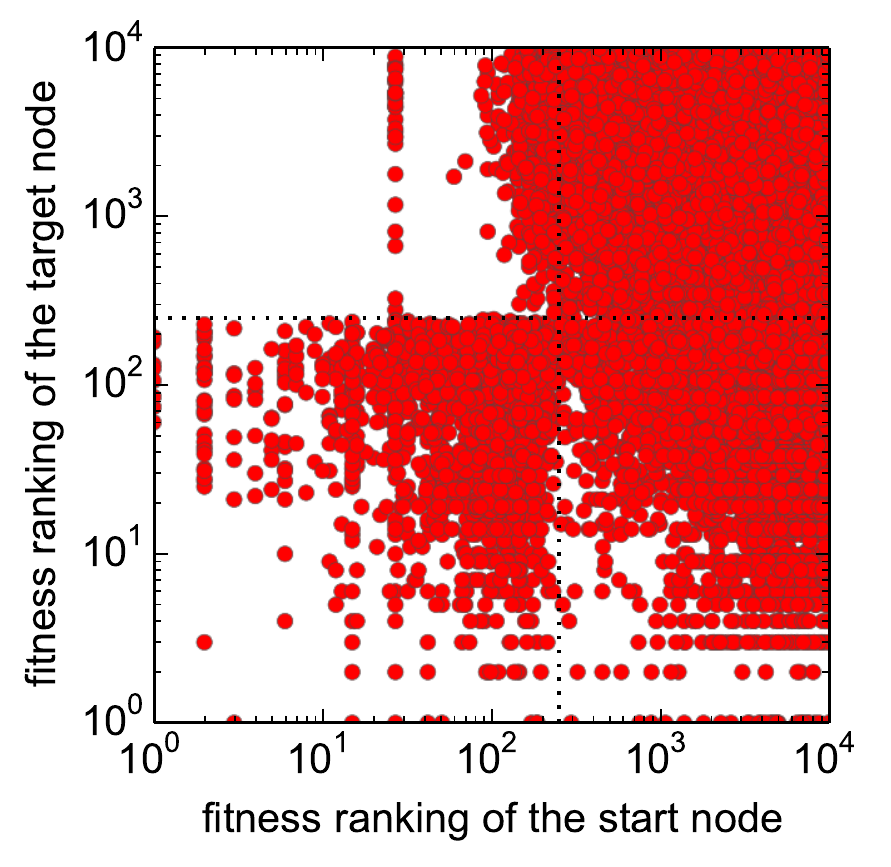}
\caption[Linking pattern in the extended fitness model (EFM)]{\textbf{Linking pattern in the extended fitness model (EFM).} Each symbol corresponds to a link from a start node whose rank according to fitness is shown on the horizontal axis to a target node whose rank according to fitness is shown on the vertical axis. The EFM parameters used to generate the network are $N=10,000$, $\theta_R=500$, $\theta_A=10000$, $H=250$. The dotted lines mark the rank position $250$ that separates high-fitness nodes (whose fitness $\eta$ is uniformly distributed in the range $[10^{-5},1]$) and low-fitness nodes (whose $\eta$ is uniformly distributed in the range $[0,10^{-5})$). We see that the EFM model produces networks where high-fitness nodes are typically pointed by other high-fitness nodes, thus creating a suitable setting for the PageRank algorithm.}
\label{SIfig:efm}
\end{figure*}

\begin{figure*}[h]
\centering
\includegraphics[scale=0.4]{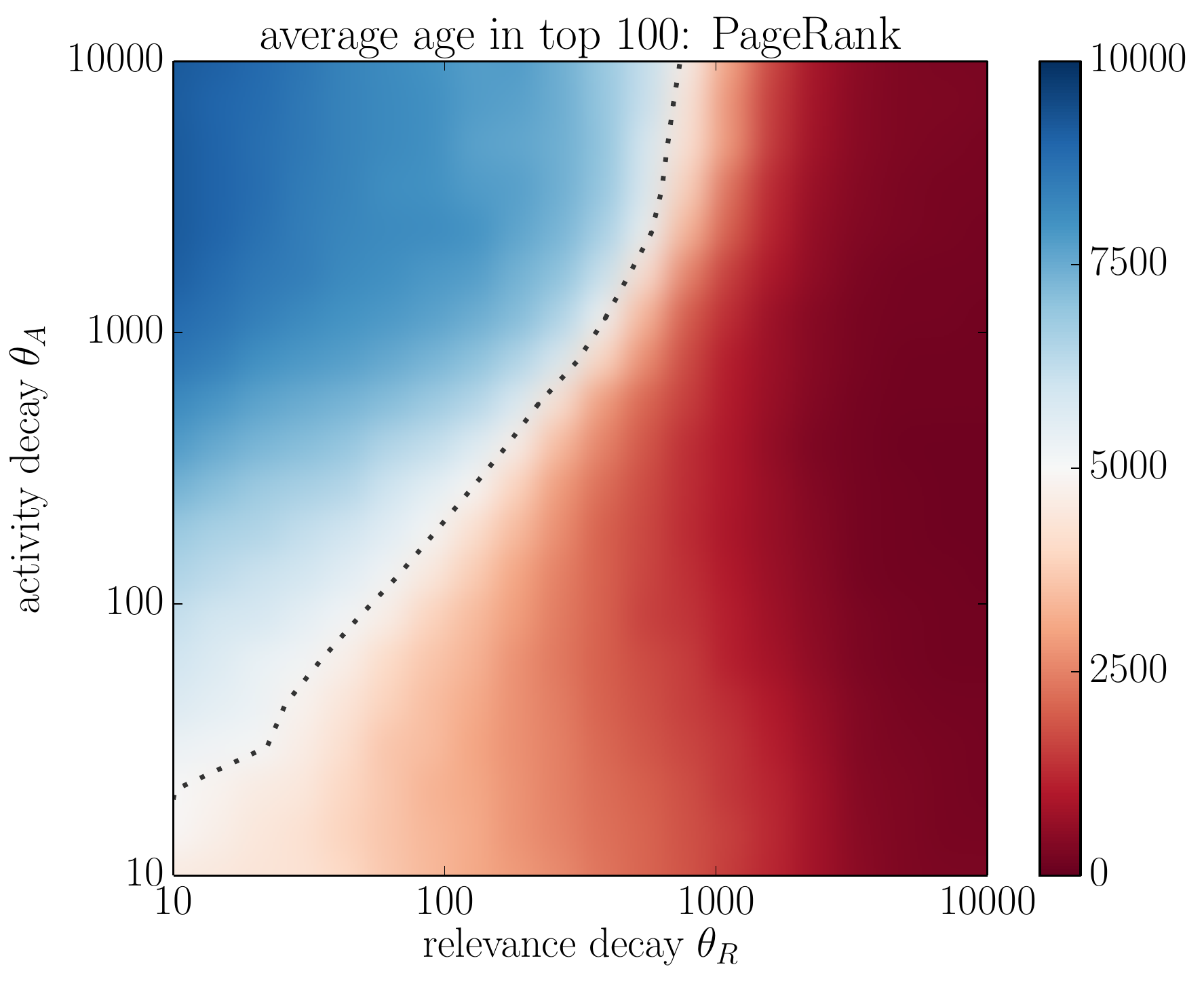}\qquad\includegraphics[scale=0.4]{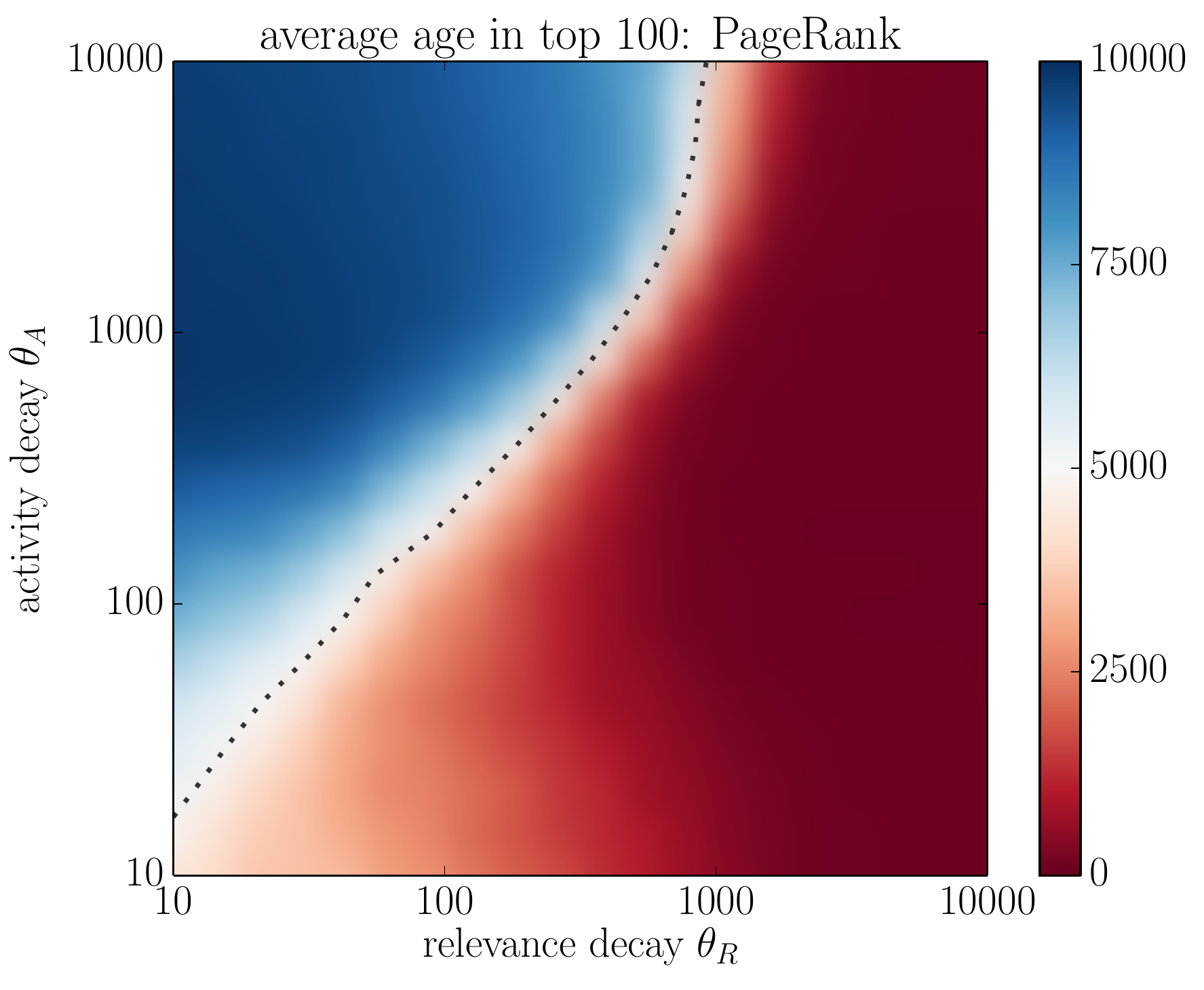}
\caption[Average birth time $\tau$ of the top $1\%$ of nodes as ranked by PageRank]{\textbf{Average birth time $\tau$ of the top $1\%$ of nodes as ranked by PageRank in the RM (left) and in the EFM (right), color online.}
When relevance decay is faster than activity decay (upper-left corners of the plots), PageRank is biased towards recent 
nodes (blue-shaded areas) in both models. When the opposite is true (lower-right corners of the plots), PageRank is biased
towards old nodes (red-shaded areas). Between the two biased regions, there is a nearly-diagonal contour (marked with the dotted line) where
the average age of top 100 nodes is $N/2=5000$ which means that the top 100 PageRank positions show no bias towards recent or old nodes.
We can conclude that PageRank is not biased only when the timescales of relevance and activity decay are in accord. Results are averaged over 6 model realizations.
}
\label{SIfig:biasmap}
\end{figure*}

\begin{figure*}[h]
\centering
\includegraphics[scale=0.4]{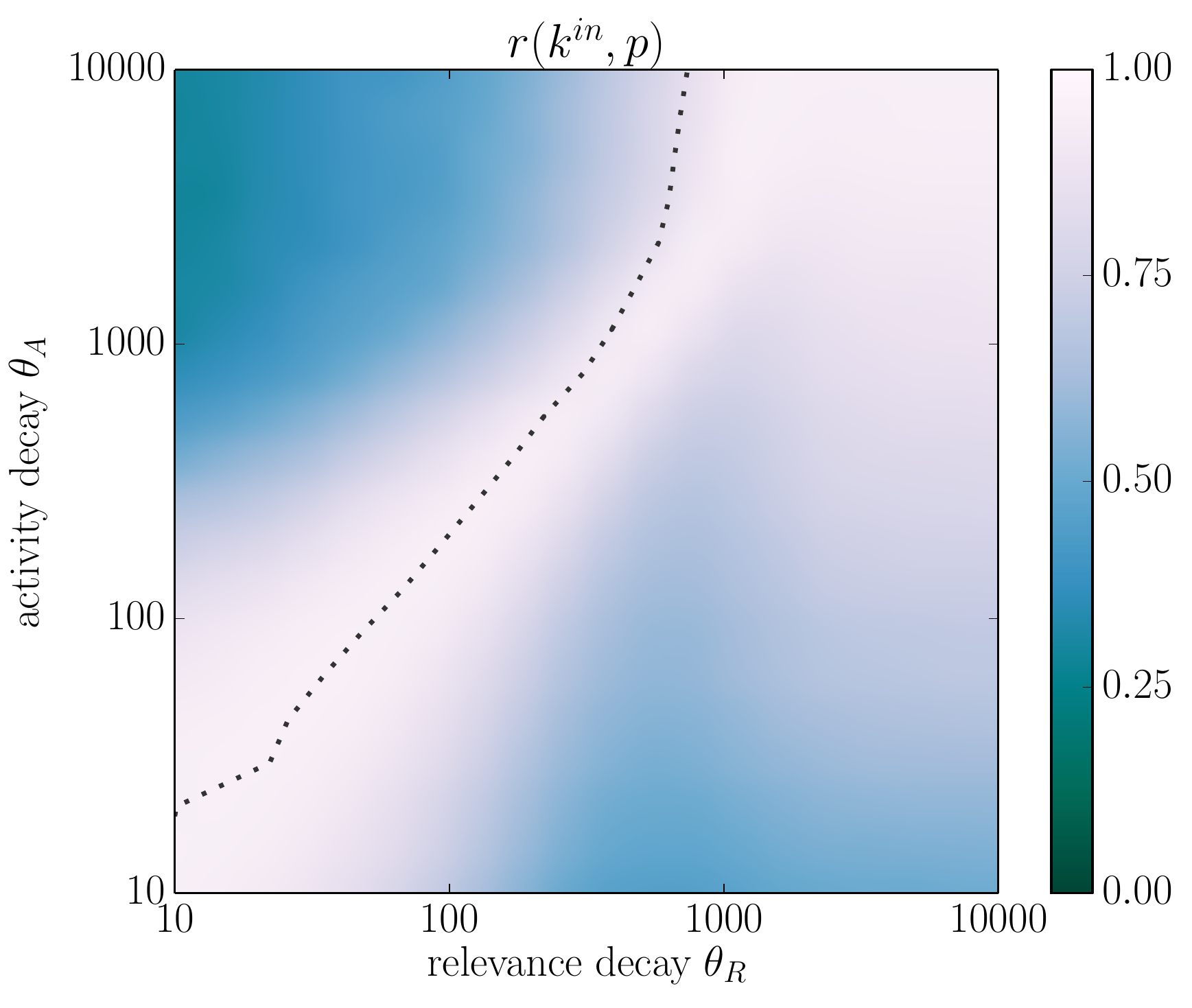}\qquad\includegraphics[scale=0.4]{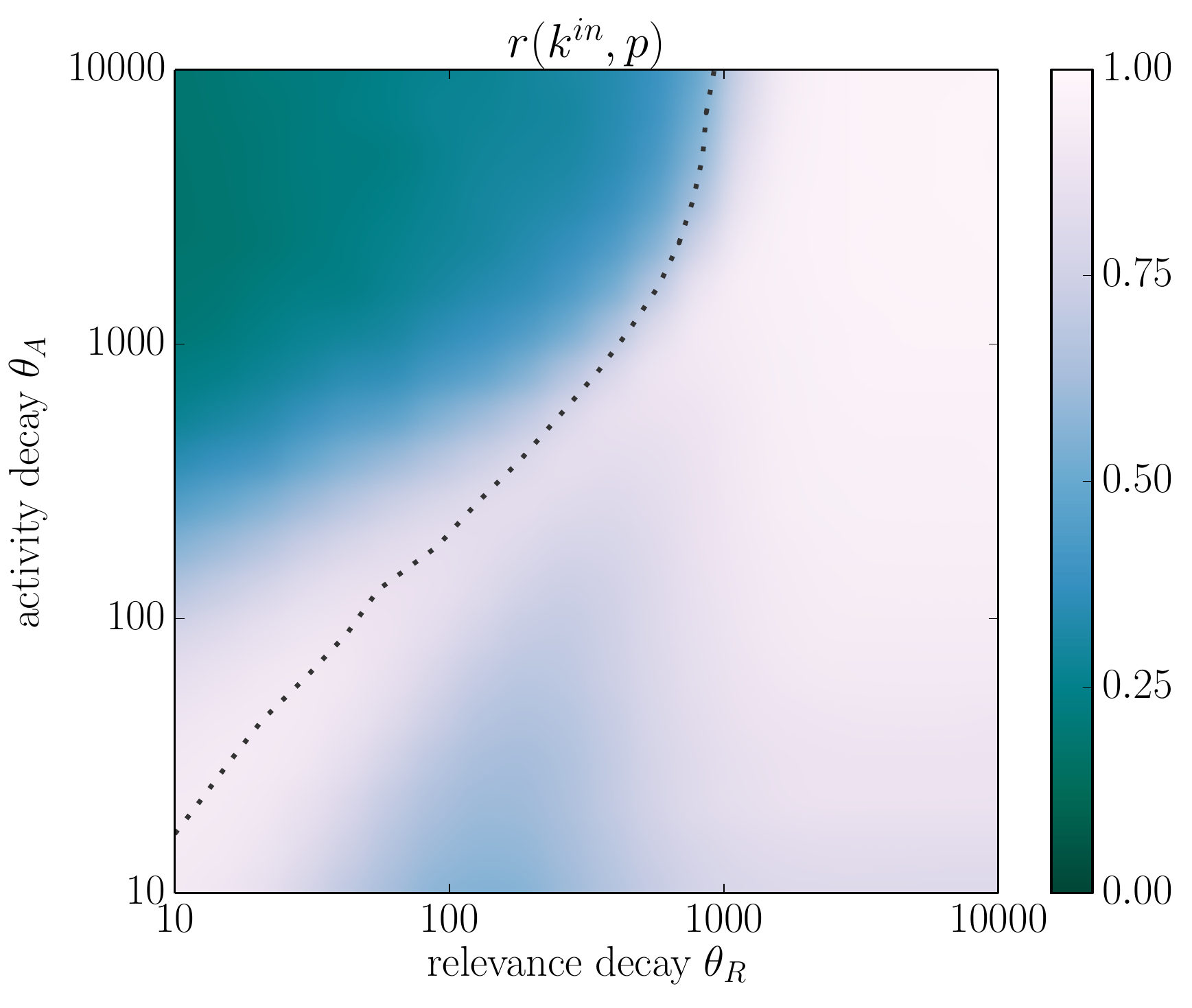}
\caption[Indegree-PageRank correlation in the RM and EFM]{\textbf{Indegree-PageRank correlation in the RM (left) and EFM (right).}
The dotted lines represent the zero-bias contours from Figure \ref{SIfig:biasmap}. When the timescales of relevance  and activity decay mismatch (upper-left and lower-right corners of the heatmaps), indegree-PageRank correlation is weak due to the time bias of PageRank. This correlation is maximal near the contour where PageRank is not biased.}
\label{SIfig:indeg_PR}
\end{figure*}

\begin{figure*}[h]
\centering
\includegraphics[scale=0.4]{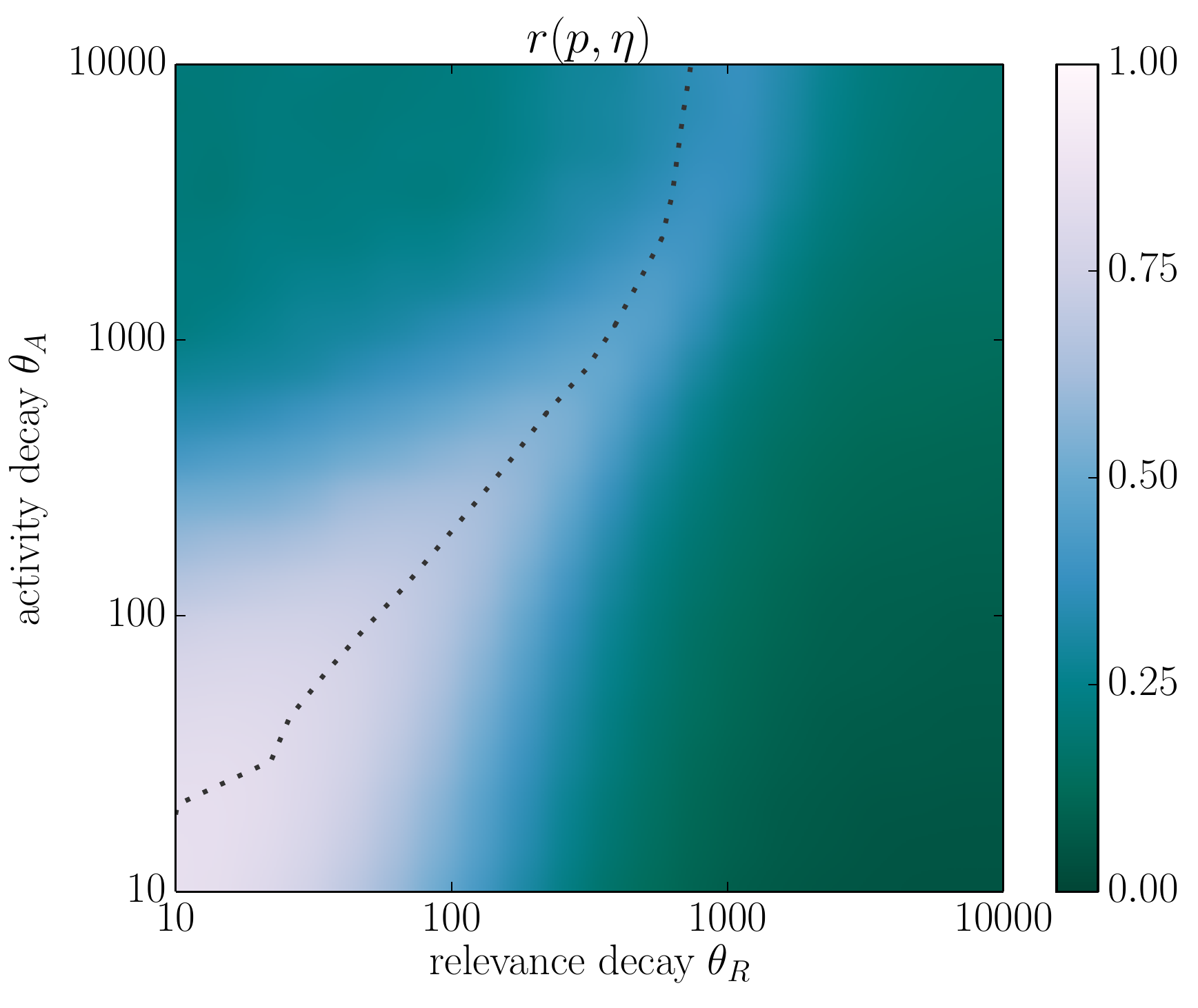}\qquad\includegraphics[scale=0.4]{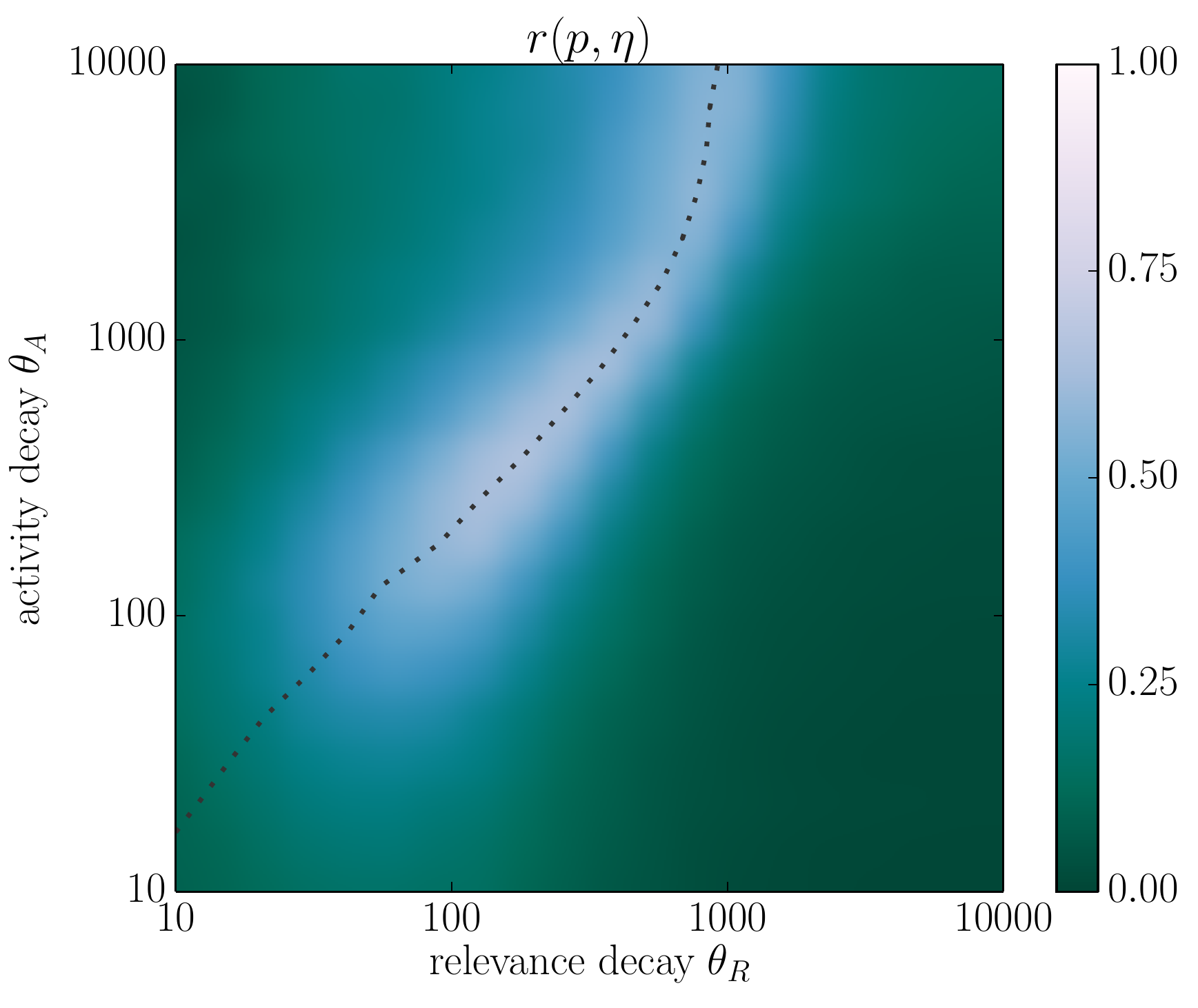}
\caption[Fitness-PageRank correlation in the RM and EFM]{\textbf{Fitness-PageRank correlation in the RM (left) and EFM (right).}
The dotted lines represent the zero-bias contours from Figure \ref{SIfig:biasmap}. When the timescales of relevance and activity decay mismatch (upper-left and lower-right corners of the heatmaps), fitness-PageRank correlation is weak due to the time bias of PageRank. PageRank performs best along the zero-bias contour.
Note that while the global maximum of PageRank's performance for the RM occurs when both $\theta_{A}$ and $\theta_{R}$ are small (lower-left corner), the global maximum for the EFM is located in the center area.
This happens because in the EFM, only a small fraction ($5\%$) of nodes are sensitive to fitness; as a result, when activity and relevance decays are too fast,
fluctuations damage the capability of indegree and PageRank to efficiently detect fitness.
By contrast, in the RM all nodes are sensitive to fitness; for this reason, $\theta_R=10$ and $\theta_A=10$ are large enough to allow the system to significantly 
perceive fitness.
}
\label{SIfig:pageRank}
\end{figure*}


\begin{figure*}[h]
\centering
\includegraphics[scale=0.4]{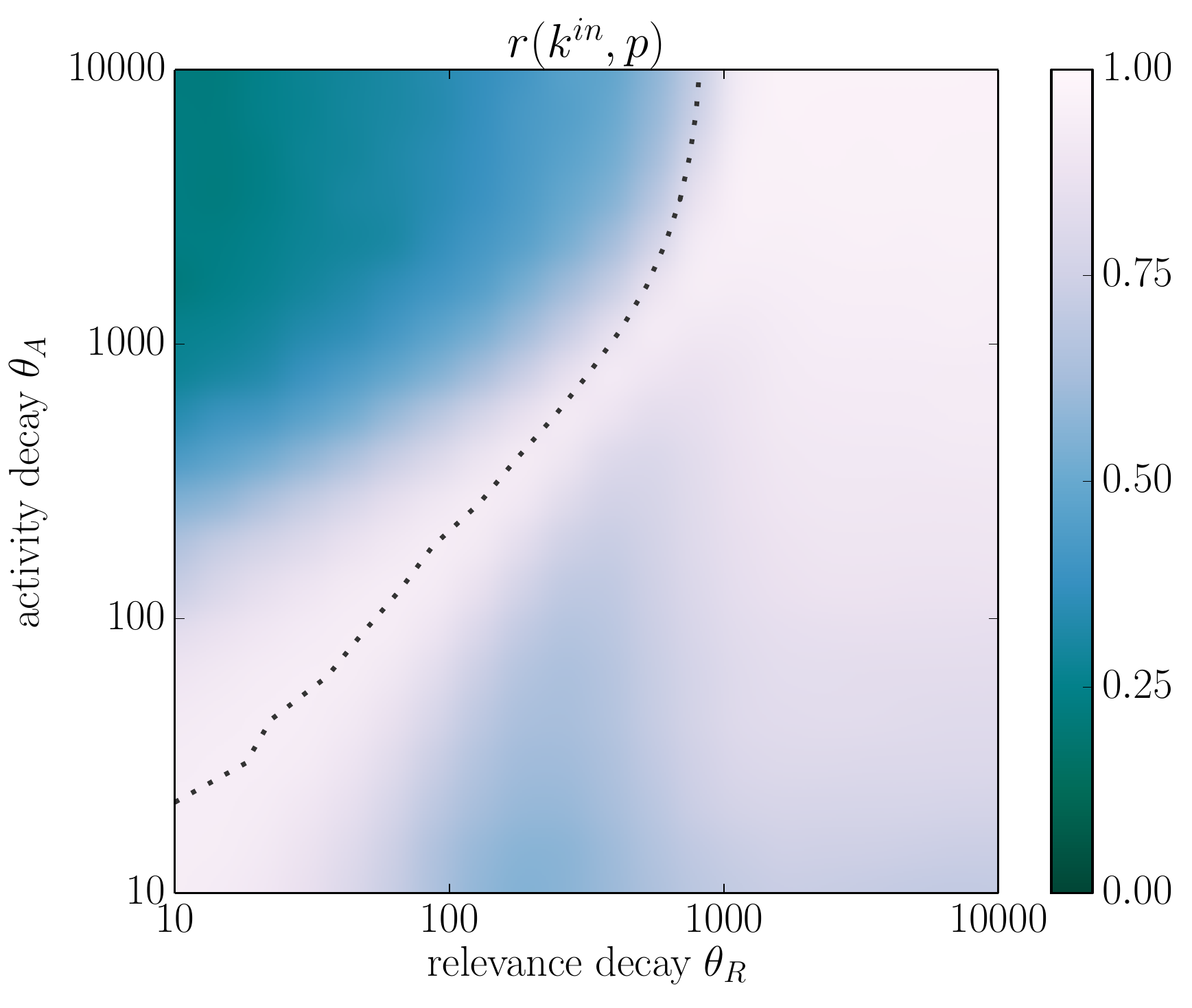}\qquad\includegraphics[scale=0.4]{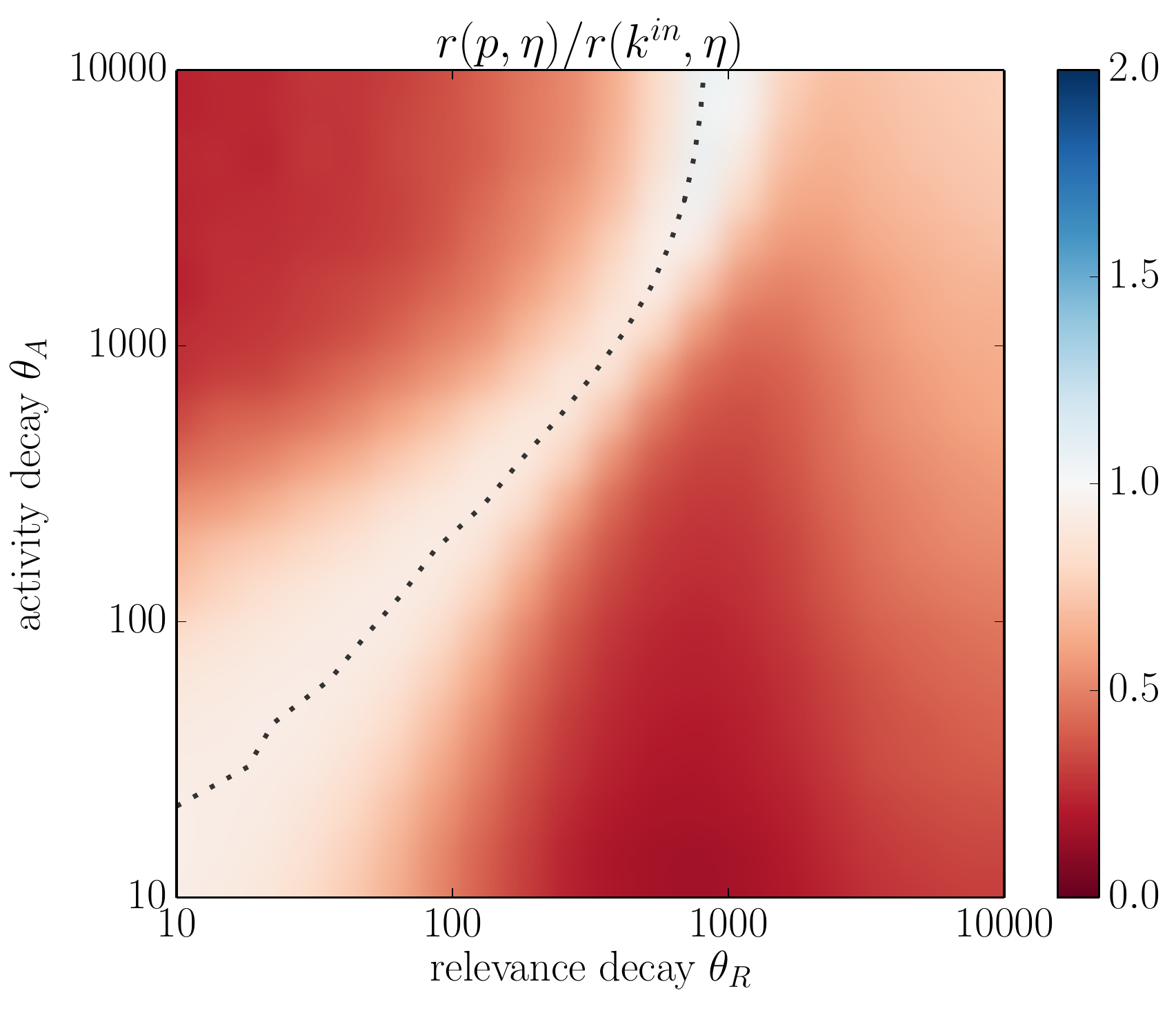}
\caption[Comparison between PageRank and indegree in the RM with uniform 
fitness distribution]{\textbf{Comparison between PageRank and indegree in the RM with 
uniform fitness distribution} (color online). Panels show the correlation $r(k^{in},p)$ (left) 
and the correlation ratio $r(p,\eta)/r(k^{in},\eta)$ (right) in the RM with node fitness $\eta$ 
distributed uniformly in $[0,1]$ (as opposed to the original exponential distribution). The dotted
lines represent the PageRank's zero-bias contour for this model (bias is again evaluated on the basis of the
average age in top 100 positions of the Pagerank ranking). Results show no significant differences from the results 
on the RM with an exponential fitness distribution (see the left panel in Figure~\ref{SIfig:indeg_PR} and the right panel in Figure~\ref{fig:RM} in the main text).}
\label{SIfig:uniform}
\end{figure*}

\begin{figure*}[h]
\centering
\includegraphics[scale=0.4]{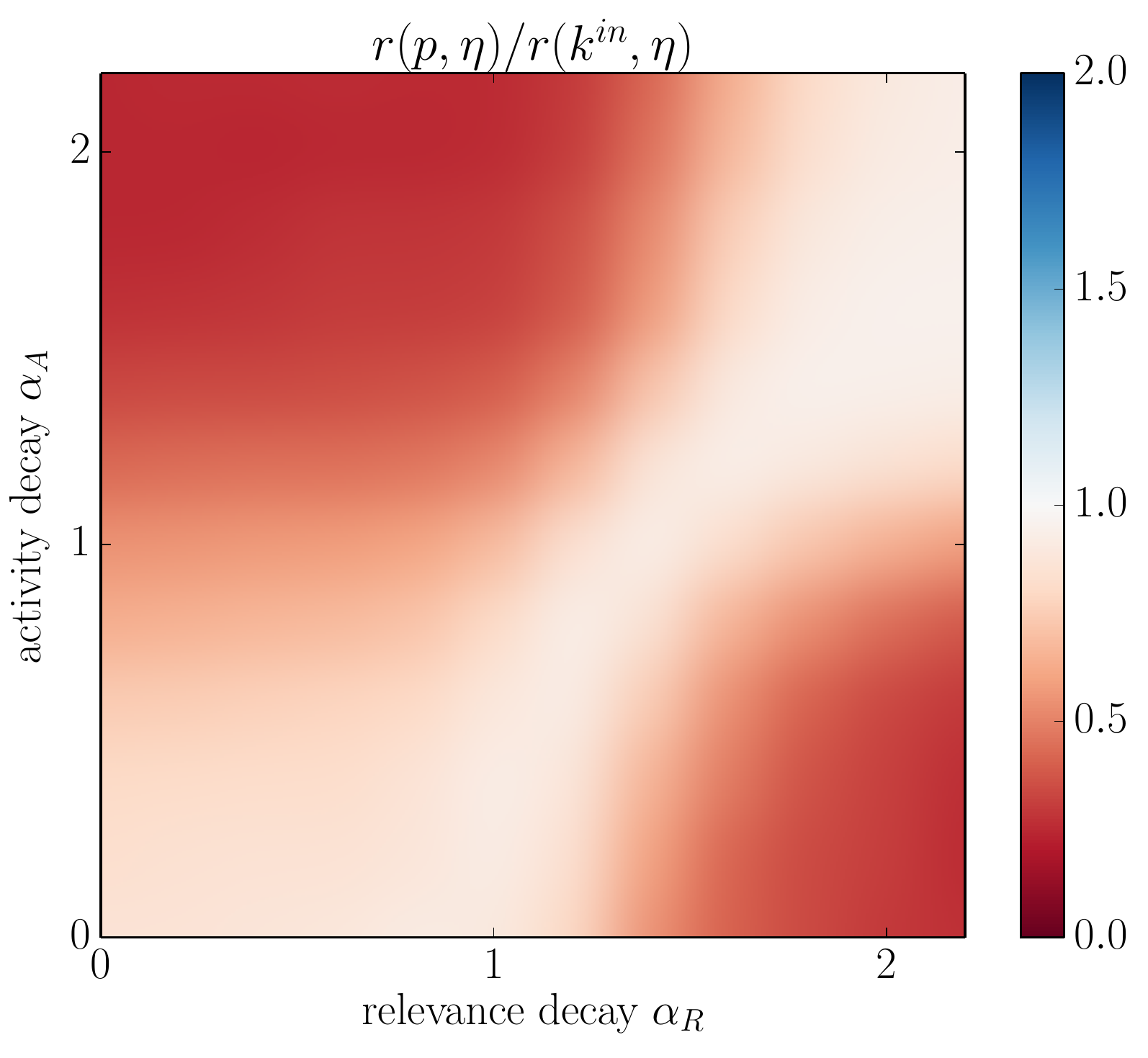}\qquad\includegraphics[scale=0.4]{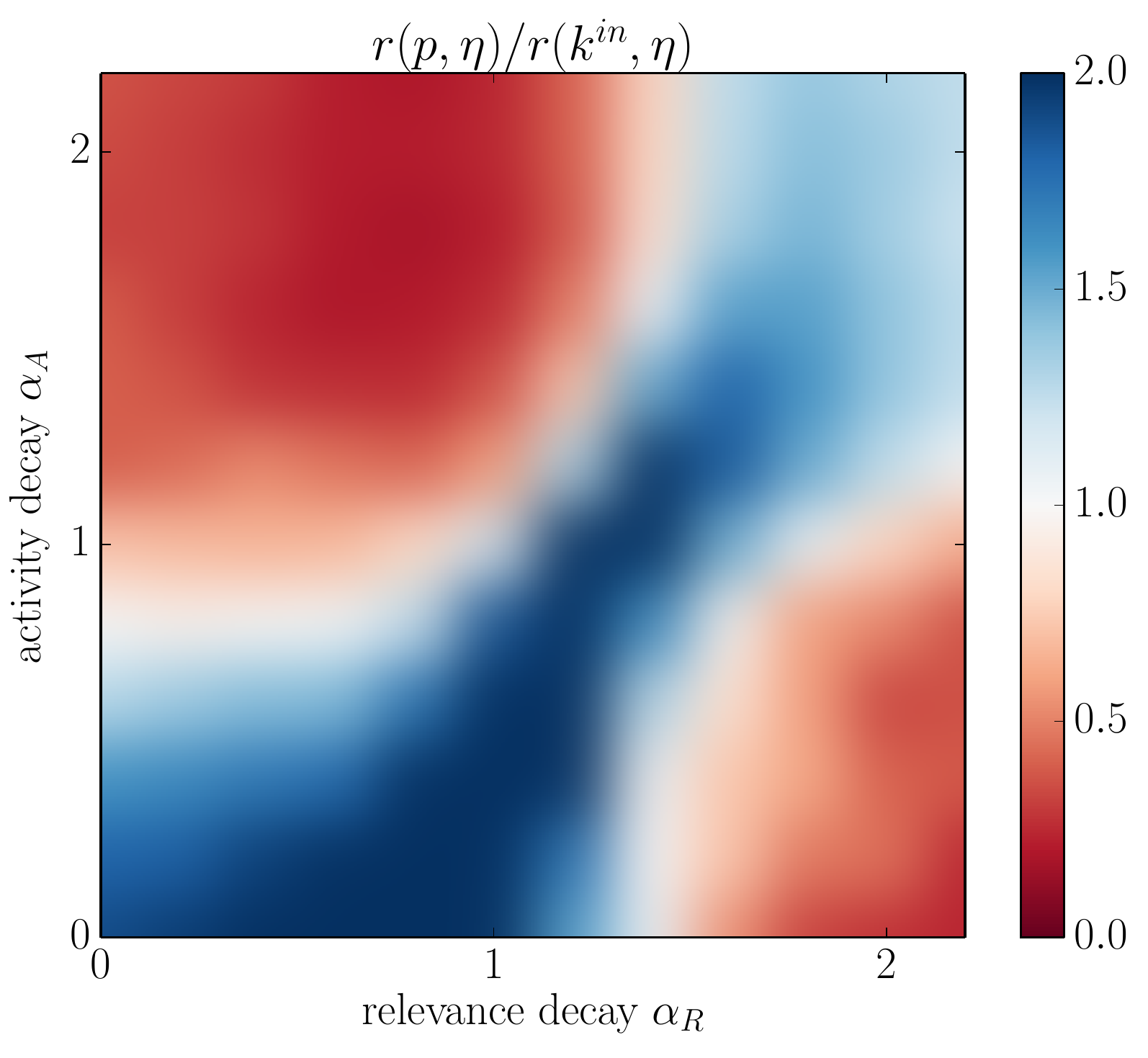}
\caption[Comparison between PageRank and indegree in the RM and EFM with power-law aging]{\textbf{Comparison
between PageRank and indegree in the RM (left) and EFM (right) with power-law aging} (color online).
We see here qualitatively the same behaviour as reported in Figures~\ref{fig:RM} and~\ref{fig:EFM} 
in the main text where exponential aging is assumed. Two regions where PageRank fails are again present: 
one in which relevance decays faster than activity (lower-right corner) and one where the opposite is true (upper-left corner).
In both the RM and the EFM, the regions where PageRank is not heavily outperformed by indegree 
include also the $(\alpha_R,\alpha_A)$ values found in Digg.com data analysis (see Table \ref{tab:fitting_digg}).
For simplicity, we focus on the parameter values 
$(\alpha_R,\alpha_A)=(1,0.4)$, even if values in Table \ref{tab:fitting_digg} are slightly different and dependent node degree group.
In the RM, PageRank performance in ranking nodes by fitness is not far from that of indegree [$(1,0.4)$ lies in the white-shaded area of left panel], 
which explains why indegree and PageRank performances in ranking nodes by relevance are close to each other in Digg.com dataset (see Fig. \ref{fig:bars} of the main text).
In the EFM, PageRank even outperforms indegree for these parameter values [$(1,0.4)$ lies in the blue-shaded area of right panel], which illustrates the main difference
between the favorable and unfavorable parameter
regions for PageRank: when PageRank is biased because of temporal effects, it fails both in the RM and in the EFM;
by contrast, when PageRank is not biased, then its performance with respect to indegree depends on the growth rule of the system
and can benefit from a suitable model such as the EFM.
}
\label{SIfig:power_law}
\end{figure*}

\begin{figure*}[h]
\centering
\includegraphics[scale=0.4]{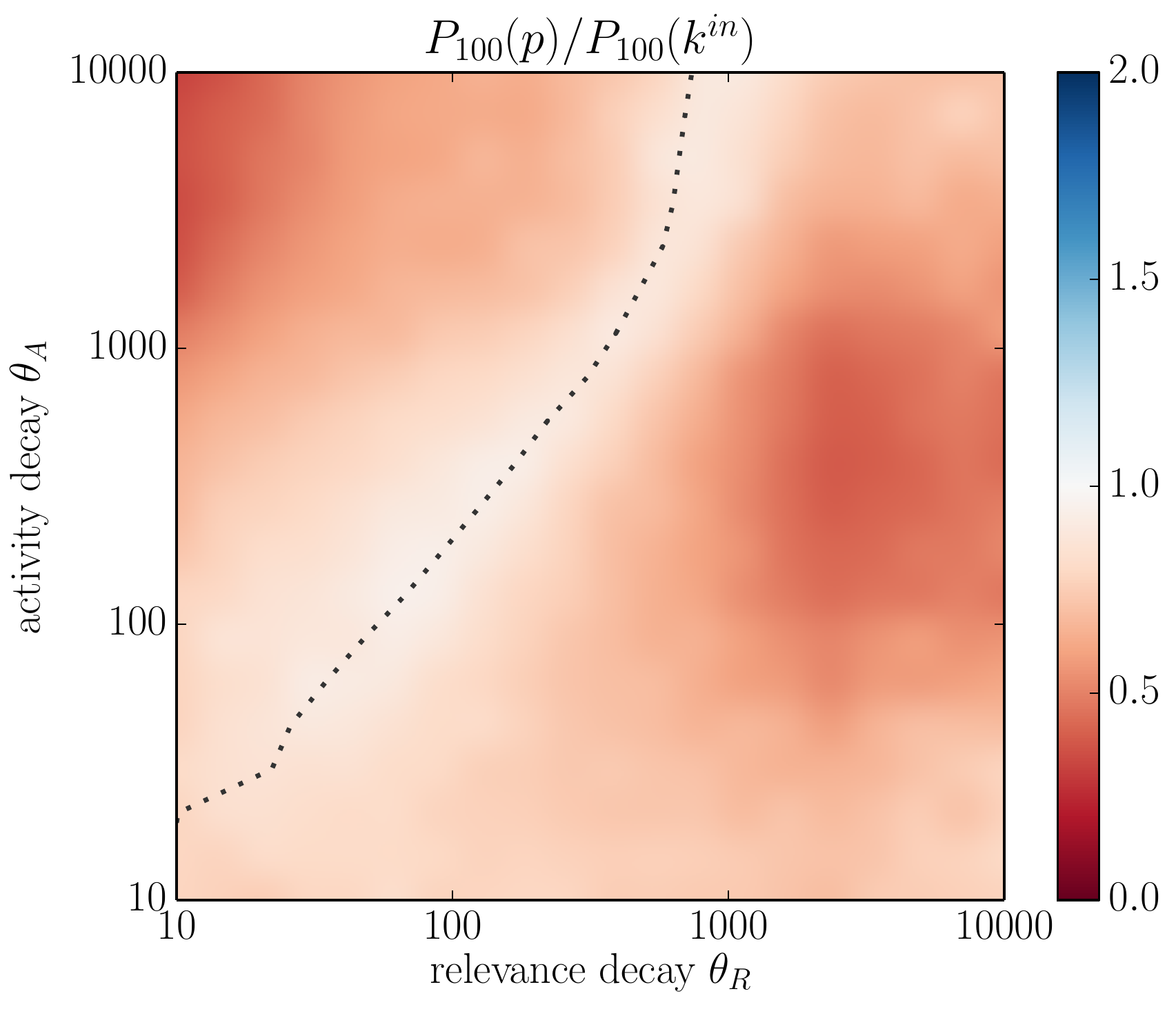}\qquad\includegraphics[scale=0.4]{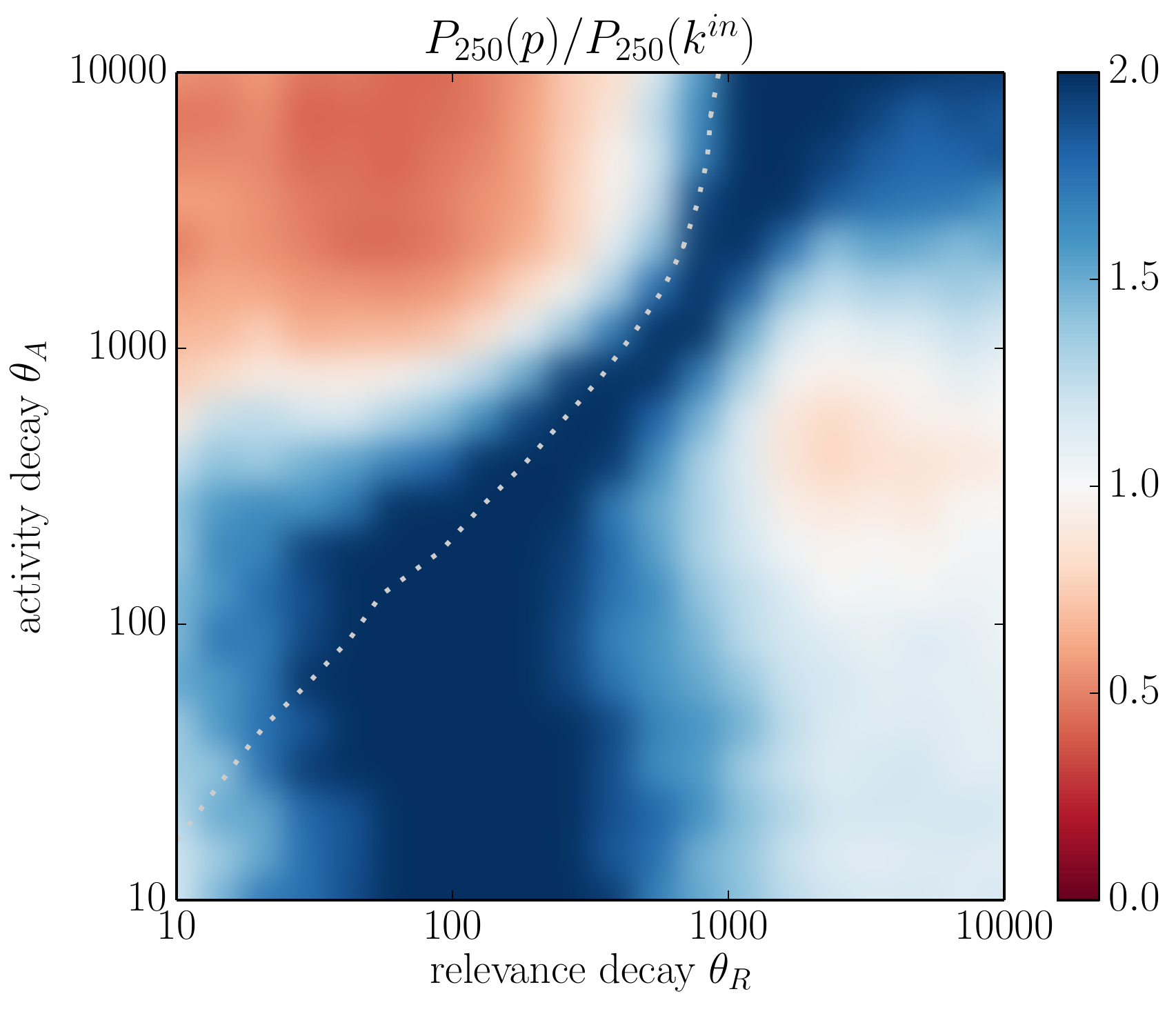}
\caption[Comparison between PageRank and indegree in the RM and EFM using precision]{\textbf{Comparison 
between PageRank and indegree in the RM (left) and EFM (right) using precision} (color online). Precision $P_X(s)$ of the ranking by 
score $s$ is the average number of nodes that are among the top $X$ places of the fitness ranking that are at the same time at the top $X$ places of the ranking by score $s$.
The dotted lines represent the zero-bias contours from Figure \ref{SIfig:biasmap}.
As for the score comparison on the basis of correlation with node fitness, PageRank's performance is again optimal 
along the contour of its zero time bias. In comparison with the correlation-based results, PageRank now lags less
behind indegree in the region where relevance decays slowly and activity decays quickly (lower-right corner). However,
both scores perform badly in this region (for the EFM, for example, $P_{250}(p) = 0.16$ in the lower-right corner as
opposed to the best-achieved precision $P_{250}(p)=0.69$ when $\theta_R=183$ and $\theta_A=263$).}
\label{SIfig:precision}
\end{figure*}

\begin{figure*}[h]
\centering
\includegraphics[scale=0.4]{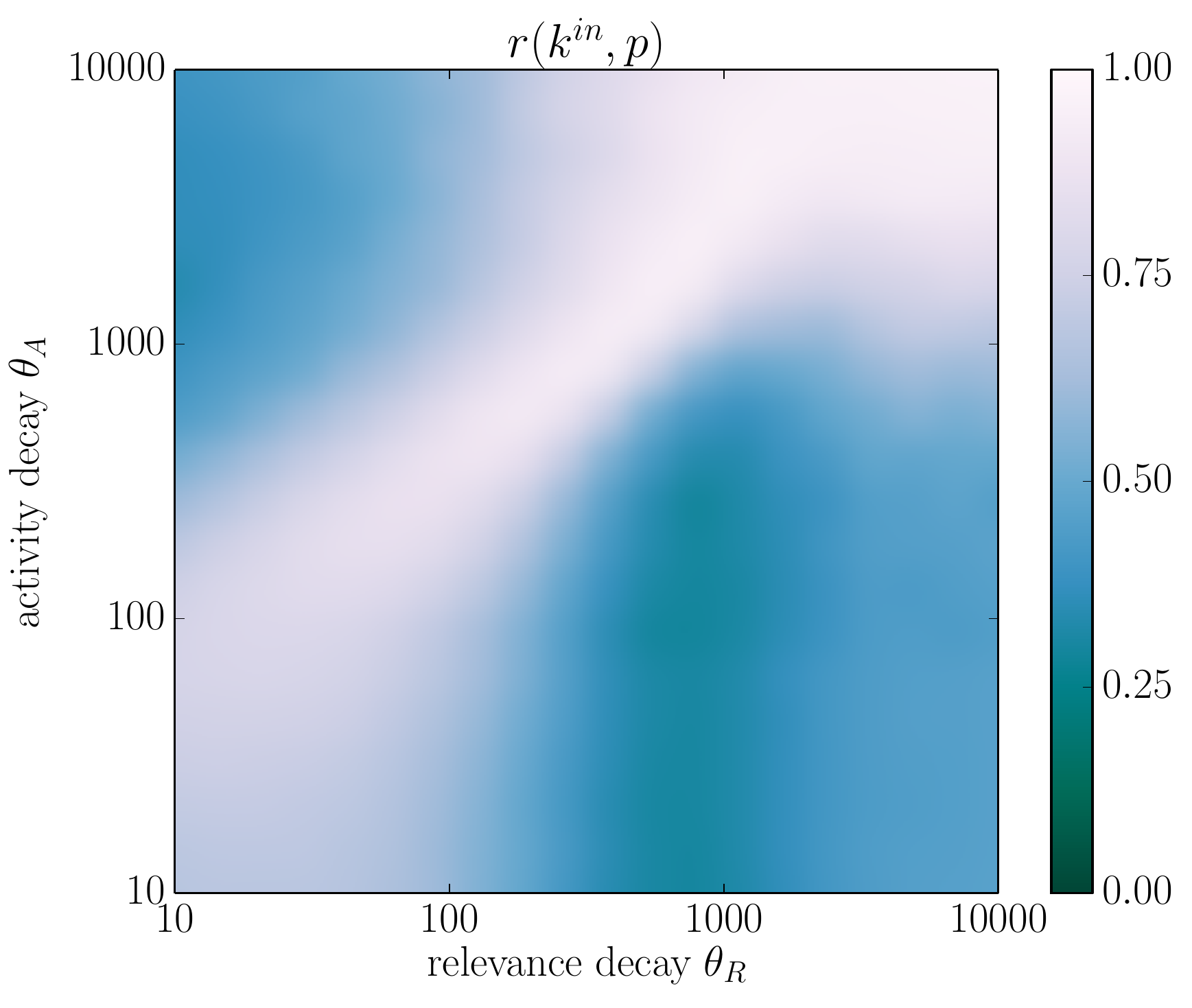}\qquad\includegraphics[scale=0.4]{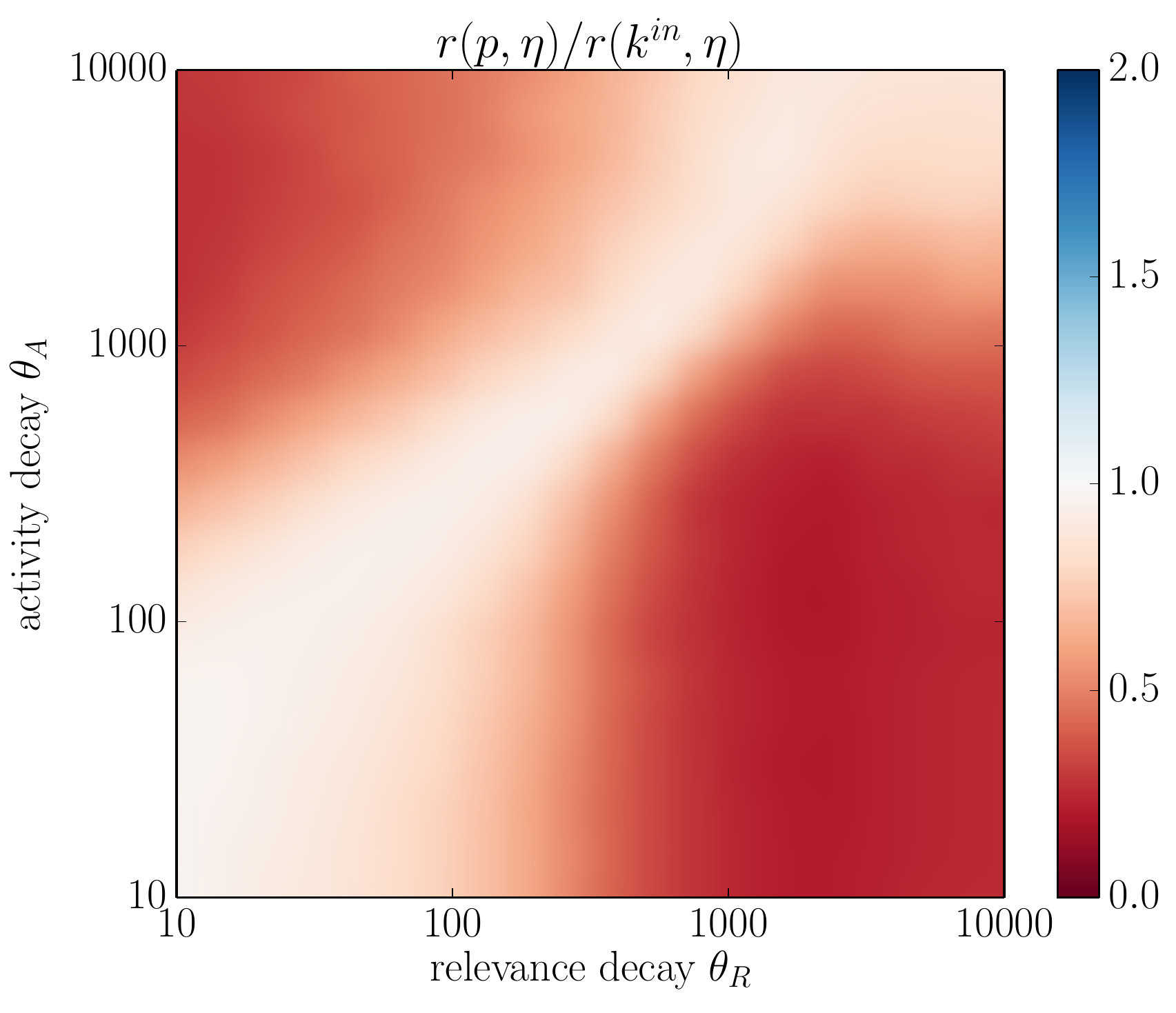}
\caption[Comparison between PageRank and indegree in the RM with accelerated growth]{\textbf{Comparison
between PageRank and indegree in the RM with accelerated growth} (color online). We grow networks according 
to the RM with the same parameters as before except that in simulation step $t$, $m(t)=20\,t/N$ existing nodes are 
sequentially chosen and create one link each. In this way, the rate at which links are created in the network grows 
linearly with network size. This aims to represent real systems where the total node activity is not normalized but
rather grows with the number of existing nodes. At the same time, the resulting number of links in the network is the
same as before when $m=10$ nodes were chosen and created one link each which makes the present model
comparable with the original one. Panels show the correlation $r(k^{in},p)$ (left) and 
the correlation ratio $r(p,\eta)/r(k^{in},\eta)$ (right). The behaviour is again qualitatively
similar to that found for uniform growth (see Figures \ref{fig:RM} and \ref{SIfig:indeg_PR}) except
that the lower-right region where PageRank fails is not even more pronounced. This further demonstrates the generality of our observations.}
\label{SIfig:accelerated}
\end{figure*}

\begin{figure}[h]
\centering
\includegraphics[scale=0.4]{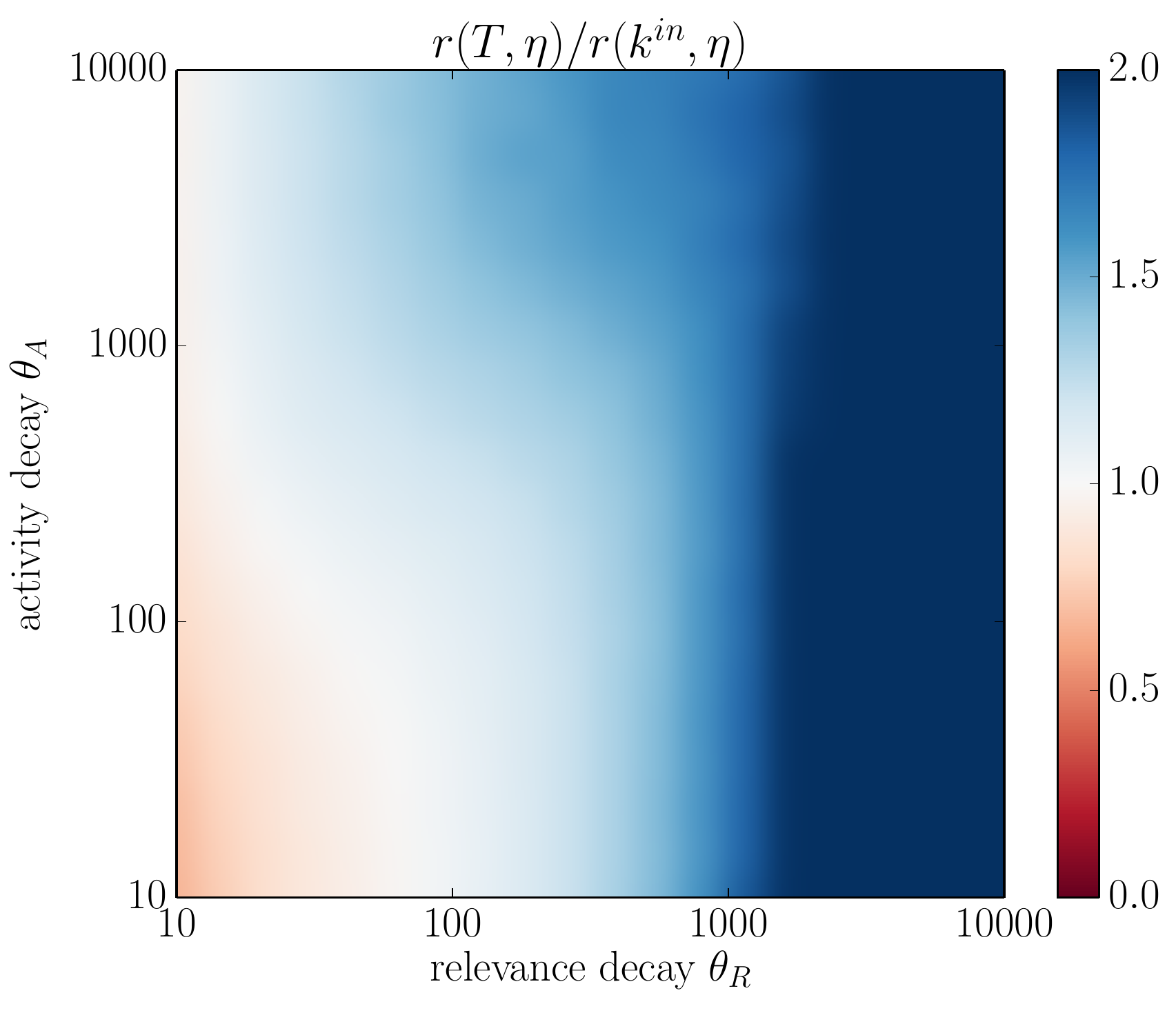}\qquad\includegraphics[scale=0.4]{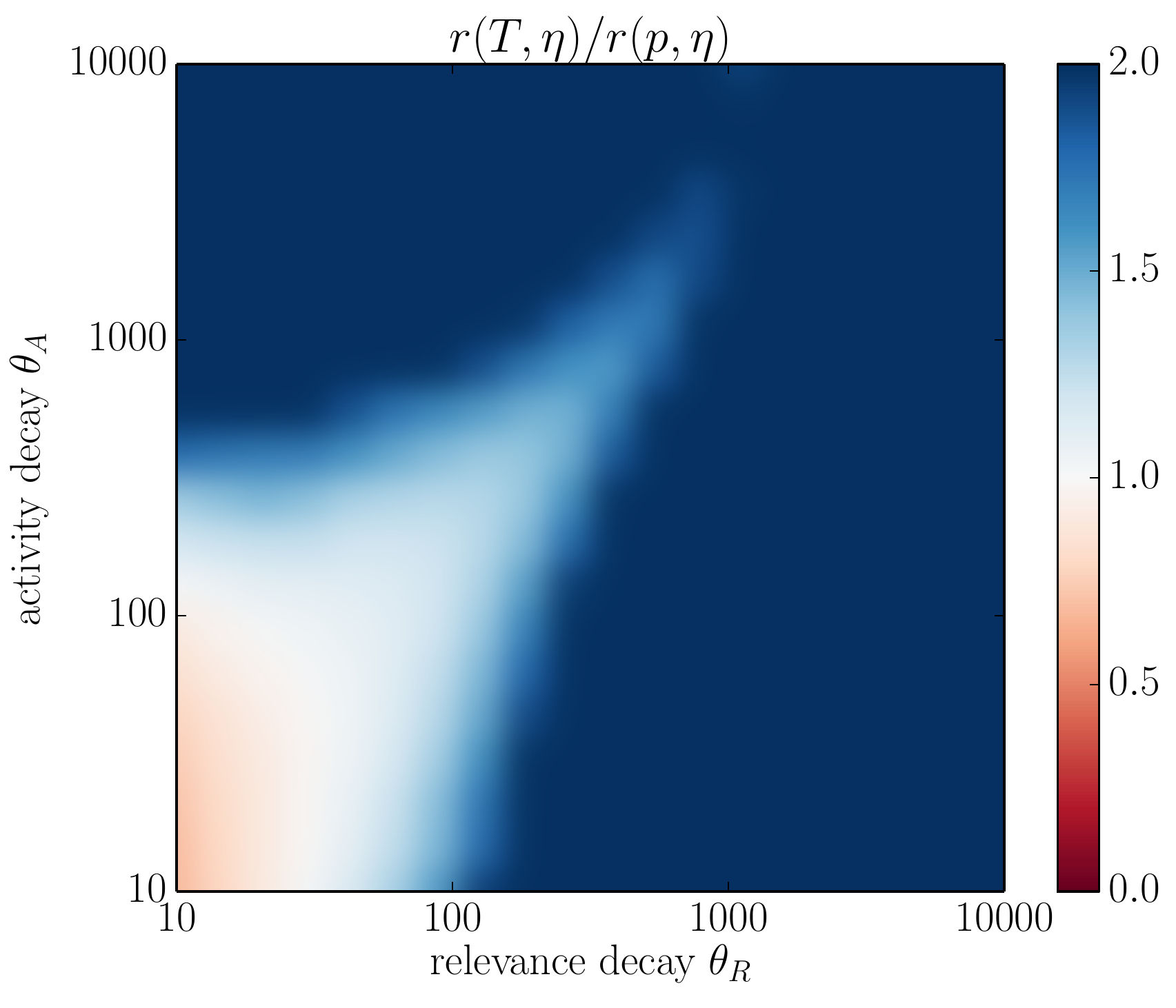}
\caption[Comparison of total relevance with indegree (left) and PageRank (right) in the RM]{\textbf{Comparison
of total relevance with indegree (left) and PageRank (right) in the RM} (color online). 
Total relevance outperforms indegree $k^{in}$ and PageRank $p$ in ranking nodes by fitness $\eta$
for a broad range of model parameter.
The performance ratios are particularly large for slow decay of relevance:
for instance, the maximum value of $r(T,\eta)/r(k^{in},\eta)$ is found for $(\theta_R,\theta_A)=(4832,14)$ [$r(T,\eta)/r(k^{in},\eta)=2.83$], and
the maximum value of $r(T,\eta)/r(p,\eta)$ is found for $(\theta_R,\theta_A)=(3359,10)$ [$r(T,\eta)/r(p,\eta)=12.49$].
These findings show that total relevance is highly informative on node fitness and might motivate the study of the ranking
by total relevance in real data.
}
\label{SIfig:tot_rel_rm}
\end{figure}

\begin{figure}[h]
\centering
\includegraphics[scale=0.4]{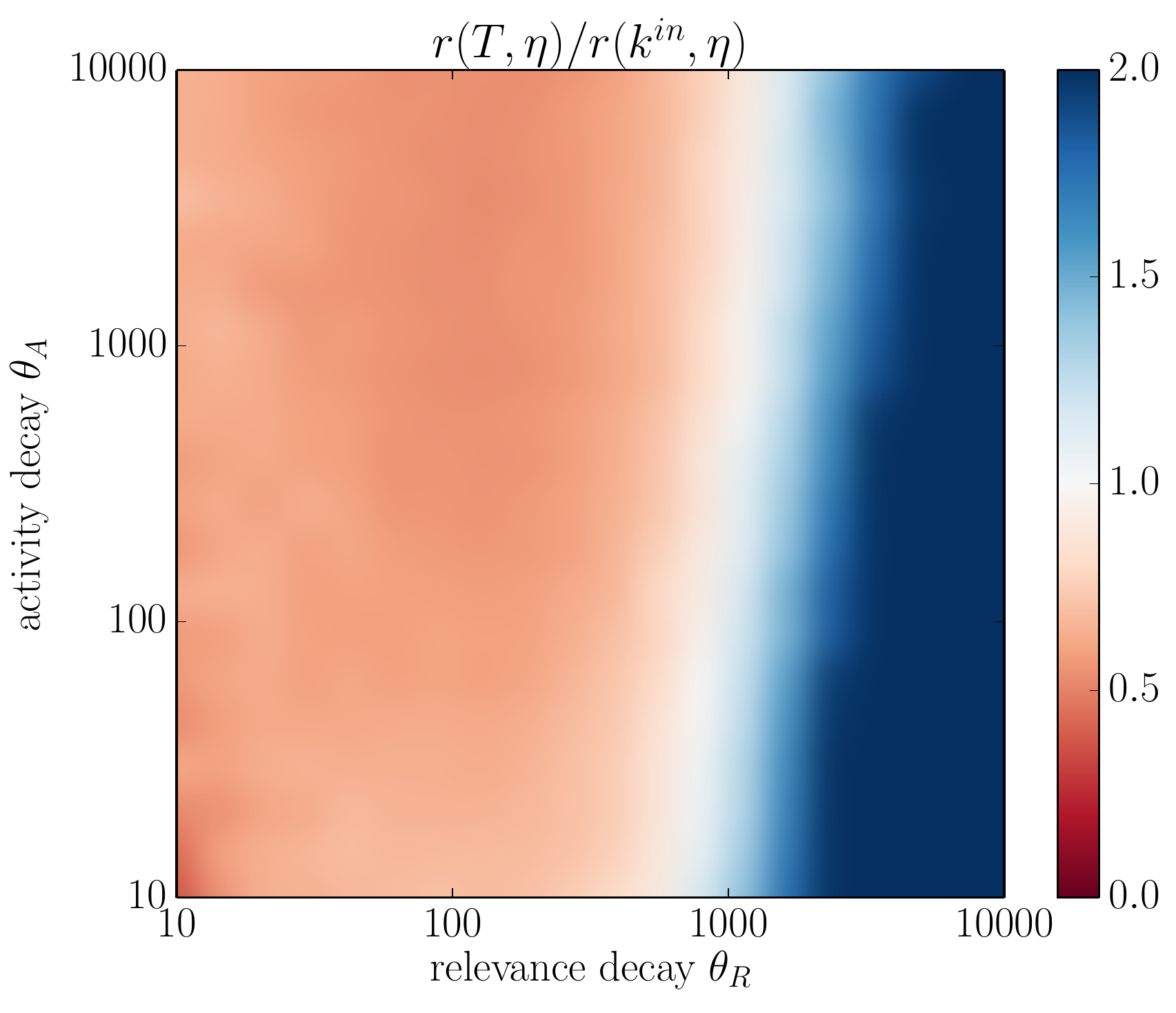}\qquad\includegraphics[scale=0.4]{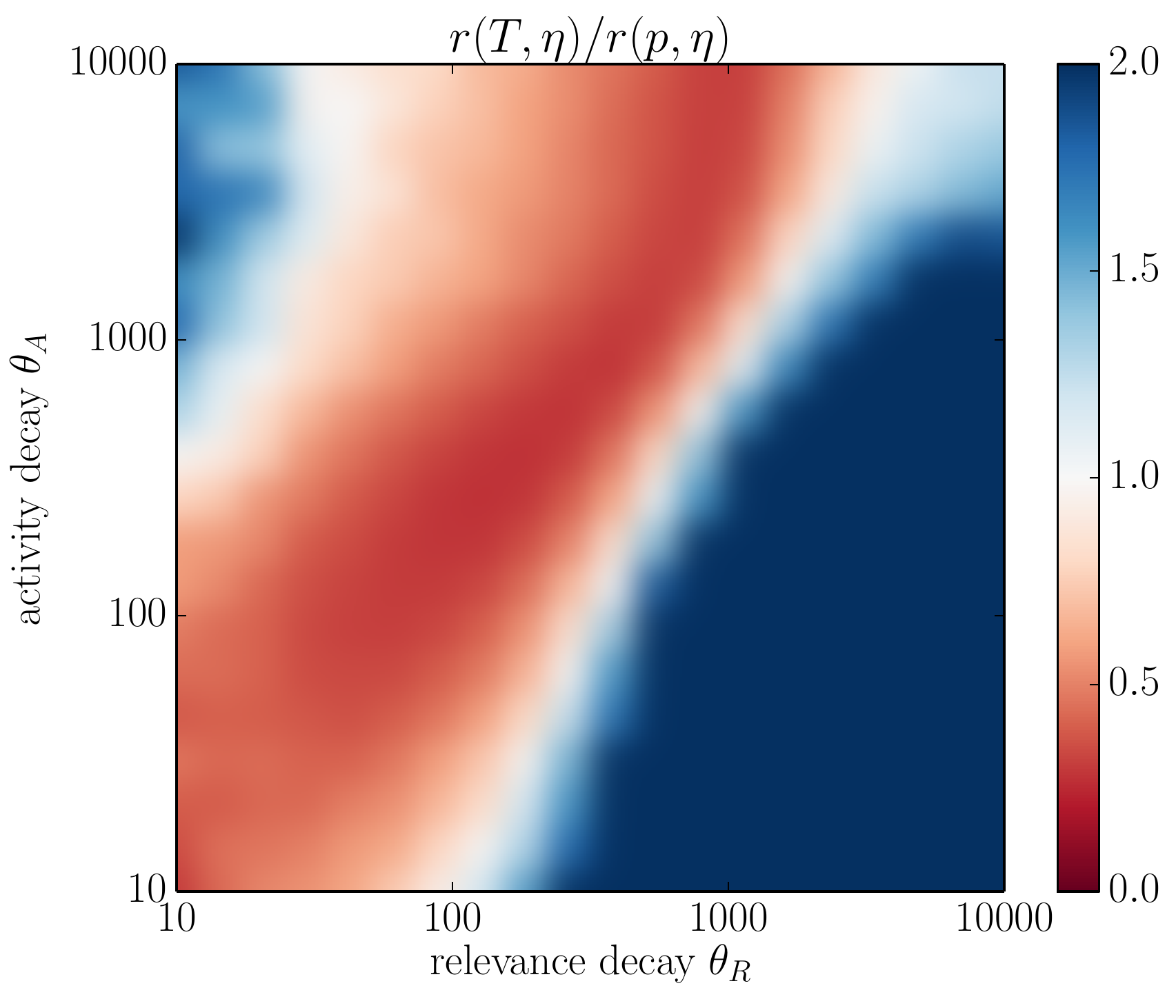}
\caption[Comparison of total relevance with indegree (left) and PageRank (right) in the EFM]{\textbf{Comparison
of total relevance with indegree (left) and PageRank (right) in the EFM} (color online).
In the data produced by the EFM, the parameter region where total relevance $T$ outperforms indegree $k^{in}$ and PageRank $p$ in ranking nodes by fitness
is smaller with respect to that observed for the RM.
The performance ratio $r(T,\eta)/r(k^{in},\eta)$ becomes large when relevance decay is slow and, as a consequence, indegree
is heavily biased towards old nodes; the maximum value of $r(T,\eta)/r(k^{in},\eta)$ is found 
for $(\theta_R,\theta_A)=(10000,10)$ [$r(T,\eta)/r(k^{in},\eta)=3.77$].
The maximum value of $r(T,\eta)/r(p,\eta)$ is found for $(\theta_R,\theta_A)=(10000,10)$ [$r(T,\eta)/r(p,\eta)=142.57$,
with $r(T,\eta)=0.172$ and $r(p,\eta)=0.001$]. On the other hand, there are broad regions of parameter values where total relevance
is outperformed by indegree and PageRank, which leaves the following question open: to what extent this failure of total relevance is due to the 
details of the model, such as the choice of $\rho(\eta)$ and the functional form of $\Pi^{in}$? Answering this question goes beyond the scope of this work.}
\label{SIfig:tot_rel_efm}
\end{figure}


\end{document}